\documentclass[aps,prl,twocolumn,10pt,amsmath,superscriptaddress,amssymb,longbibliography]{revtex4-2}
\usepackage{epsfig,epstopdf,dcolumn,amsbsy,bm,color,CJK,amsmath,amssymb,amsfonts,appendix}
\usepackage{graphicx,algorithm,enumerate,makeidx,braket}
\usepackage[usenames,dvipsnames]{xcolor}
\usepackage[driverfallback=dvipdfmx,colorlinks,linkcolor=blue,citecolor=blue,urlcolor=blue]{hyperref}
\usepackage{pifont,textcomp,helvet}

\begin{document}

\title{Magnetic-Field-Calibration-Free Determination of the Hyperfine Constant $A$ in Ultracold Fermi gases of $^{40}$K}

\author{Yajing Yang}
\thanks{These authors contributed equally to this work.}
\affiliation{State Key Laboratory of Quantum Optics Technologies and Devices, \\  Institute of Opto-electronics, Collaborative Innovation Center of Extreme Optics, Shanxi University, Taiyuan, Shanxi 030006, China}
\author{Biao Shan}
\thanks{These authors contributed equally to this work.}
\affiliation{State Key Laboratory of Quantum Optics Technologies and Devices, \\  Institute of Opto-electronics, Collaborative Innovation Center of Extreme Optics, Shanxi University, Taiyuan, Shanxi 030006, China}
\author{Yuhang Zhao}
\affiliation{State Key Laboratory of Quantum Optics Technologies and Devices, \\  Institute of Opto-electronics, Collaborative Innovation Center of Extreme Optics, Shanxi University, Taiyuan, Shanxi 030006, China}
\author{Jiahui Shen}
\affiliation{State Key Laboratory of Quantum Optics Technologies and Devices, \\  Institute of Opto-electronics, Collaborative Innovation Center of Extreme Optics, Shanxi University, Taiyuan, Shanxi 030006, China}
\author{Zhuxiong Ye}
\affiliation{State Key Laboratory of Quantum Optics Technologies and Devices, \\  Institute of Opto-electronics, Collaborative Innovation Center of Extreme Optics, Shanxi University, Taiyuan, Shanxi 030006, China}
\author{Liangchao Chen}
\affiliation{State Key Laboratory of Quantum Optics Technologies and Devices, \\  Institute of Opto-electronics, Collaborative Innovation Center of Extreme Optics, Shanxi University, Taiyuan, Shanxi 030006, China}
\affiliation{Hefei National Laboratory, Hefei, Anhui 230088, China.}
\author{Zengming Meng}
\affiliation{State Key Laboratory of Quantum Optics Technologies and Devices, \\  Institute of Opto-electronics, Collaborative Innovation Center of Extreme Optics, Shanxi University, Taiyuan, Shanxi 030006, China}
\affiliation{Hefei National Laboratory, Hefei, Anhui 230088, China.}
\author{Pengjun Wang}
\affiliation{State Key Laboratory of Quantum Optics Technologies and Devices, \\  Institute of Opto-electronics, Collaborative Innovation Center of Extreme Optics, Shanxi University, Taiyuan, Shanxi 030006, China}
\affiliation{Hefei National Laboratory, Hefei, Anhui 230088, China.}
\author{Wei Han}
\email[Contact author:]{hanwei.irain@gmail.com}
\affiliation{State Key Laboratory of Quantum Optics Technologies and Devices, \\  Institute of Opto-electronics, Collaborative Innovation Center of Extreme Optics, Shanxi University, Taiyuan, Shanxi 030006, China}
\affiliation{Hefei National Laboratory, Hefei, Anhui 230088, China.}
\author{Jing Zhang}
\email[Contact author:]{jzhang74@sxu.edu.cn}
\affiliation{State Key Laboratory of Quantum Optics Technologies and Devices, \\  Institute of Opto-electronics, Collaborative Innovation Center of Extreme Optics, Shanxi University, Taiyuan, Shanxi 030006, China}
\affiliation{Hefei National Laboratory, Hefei, Anhui 230088, China.}
\author{Lianghui Huang}
\email[Contact author:]{huanglh06@sxu.edu.cn}
\affiliation{State Key Laboratory of Quantum Optics Technologies and Devices, \\  Institute of Opto-electronics, Collaborative Innovation Center of Extreme Optics, Shanxi University, Taiyuan, Shanxi 030006, China}
\affiliation{Hefei National Laboratory, Hefei, Anhui 230088, China.}

\date{\today }

\begin{abstract}
Hyperfine constant $A$ is a key parameter of the hyperfine structure and underpins precision spectroscopy and metrology. In this Letter, we develop a magnetic-field-calibration-free method for determining the ground-state hyperfine constant $A$ in an ultracold $^{40}$K Fermi gas by utilizing a pair of magnetically insensitive (“clock”) transitions. This overcomes the stringent magnetic-field calibration requirements of conventional methods. We measure the transition frequency between these two magnetically insensitive transitions with Hz-level resolution over a range of magnetic fields, and obtain the ground-state hyperfine constant $A = -h\times 285.730536(2)\,\mathrm{MHz}$, corresponding to an absolute uncertainty of about $2\,\mathrm{Hz}$. Our value reduces the uncertainty by nearly three orders of magnitude compared with previous determinations, providing a substantially improved reference for high-precision spectroscopy and metrology with $^{40}$K.
\end{abstract}


\maketitle

The ground-state hyperfine structure constant $A$ is the key atomic-structure parameter characterizing the ground-state hyperfine splitting. With the development of atomic beam~\cite{Ramsey1956Beams}, atomic fountain~\cite{Metcalf1999Fountain},  and optical frequency-comb techniques~\cite{Hansch2002Nature}, the ground-state hyperfine constants $A$ of alkali isotopes such as $^{6/7}\mathrm{Li}$~\cite{Schlecht1966LiHFS}, $^{23}\mathrm{Na}$~\cite{Arditi1958NaHFS}, $^{39/41}\mathrm{K}$~\cite{beckmann1974ZFP, Peper2019K39GroundHFS}, $^{85}\mathrm{Rb}$~\cite{Bender1958Rb87HFS, Yuzhu2019PRA}, $^{87}\mathrm{Rb}$~\cite{ovchinnikov2015Metrologia}, and $^{133}\mathrm{Cs}$~\cite{Essen1955CsClock, Parry1958PRL} have been determined with extremely high accuracy, so that the last significant digit typically corresponds to a frequency uncertainty better than the Hz level~\cite{Arimondo1977RMP, Allegrini2022JPC}. In particular, using atomic fountain and atomic-beam resonance techniques, which operate on a magnetically insensitive clock transition under magnetic shielding and stringent magnetic-field calibration, the ground-state hyperfine splittings of $^{87}$Rb and $^{133}$Cs have been measured with uncertainties of a few $\mathrm{\mu Hz}$ for $^{87}$Rb~\cite{ovchinnikov2015Metrologia} and about $0.1\,\mathrm{Hz}$ for $^{133}$Cs~\cite{Parry1958PRL}, respectively.

However, not all alkali isotopes have reached this level of precision.
For the fermionic isotope $^{40}$K, traditional methods face significant difficulties in producing high-density atomic samples, primarily due to $^{40}$K’s extremely low natural abundance of only about 0.0117\%~\cite{Wei2007Enriched40K, Prohaska2016PAC}. Consequently, performing high-resolution absorption spectroscopy to determine the ground-state hyperfine constant $A$ of $^{40}$K is intrinsically challenging. The early measurement of $A$ was determined using atomic beam spectroscopy at high magnetic field and without magnetic shielding~\cite{Zacharias1942PR, Davis1949PR, Eisinger1952PR}.
The corresponding value, $A$$=$$-h\times285.7308(24)\,\mathrm{MHz}$, obtained from a later evaluation of those measurements~\cite{Arimondo1977RMP}, has been widely used in early studies of laser cooling, magneto-optical trapping, and degenerate Fermi gases~\cite{Falke2006PRA, Behrle2011PRA, Hanley2015JOSAB}. However, with the development of ultracold-atom experiments, its limited accuracy has now become a constraint. For example, it affects magnetic-field calibration, Feshbach-resonance fitting, and the extraction of spin-dependent interaction strengths, making magnetic-field scales difficult to compare across experiments and preventing further reduction of systematic uncertainties~\cite{Julienne2008PRL,Zwierlein2012PRA}.

A more precise determination of the ground-state hyperfine constant $A$ for $^{40}\mathrm{K}$ is therefore crucial. In this Letter, we develop a magnetic-field-calibration-free experimental scheme for determining the ground-state hyperfine constant $A$ through high-precision measurements of a pair of magnetically insensitive hyperfine transitions. Unlike conventional methods, this approach operates in an ultracold, quantum-degenerate regime in an optical trap, which greatly enhances the atomic density, and it exploits the magnetic-field insensitivity of the two transitions, thereby allowing the experiment to proceed without magnetic shielding and magnetic-field calibration. Using the developed method, we determined the ground-state hyperfine constant of ${}^{40}\mathrm{K}$ to be $A = -h\times 285.730536(2)\,\mathrm{MHz}$ with an absolute uncertainty of about $2~\mathrm{Hz}$, reducing the uncertainty by nearly three orders of magnitude compared with previous determinations~\cite{Zacharias1942PR, Davis1949PR, Eisinger1952PR, Arimondo1977RMP}.

\begin{figure}[t]
\centering
\includegraphics[width=3.2in]{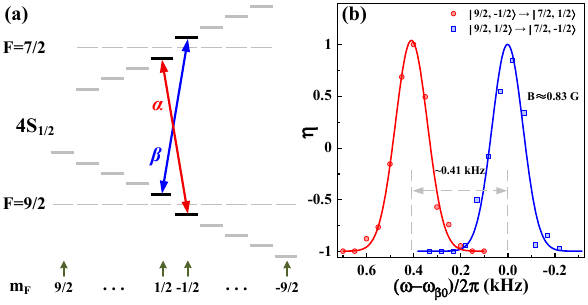}
\caption{(a)~Energy-level diagram showing the single-photon microwave magnetically insensitive transitions $\alpha$ and $\beta$ in $^{40}$K. (b)~Microwave spectra of the two sets of magnetically insensitive transitions $\alpha$ and $\beta$. The measurements are performed using a Gaussian-shaped $\pi$ pulse with a duration of 10.5 ms and optimized amplitude under a bias magnetic field of 0.83~G. Here, $\omega_{\beta0}$ denotes the resonance frequency for transition $\beta$.
}
\label{setup}
\end{figure}

We consider two magnetically insensitive single-photon transitions~\cite{Shan2026PRA} labeled $\alpha$ for $|9/2, -1/2\rangle$$\Leftrightarrow$$|7/2$$,$$1/2\rangle$ and $\beta$ for $|9/2, 1/2\rangle$$\Leftrightarrow$$|7/2$$,$$-1/2\rangle$ within the $^{40}\mathrm{K}$ ground-state hyperfine manifold, as shown in Fig.~\ref{setup}(a).
In the presence of an external magnetic field $B$, the Zeeman energies of the hyperfine states depend on the magnetic quantum number $m_{F}$ and are described by the Breit--Rabi formula~\cite{Rabi1931PR,Daniel2019K40}
\begin{equation}\label{Breit}
\begin{split}
E(B)=&-\frac{A}{4}+\mu_B g_{I}m_{F} B \\ & \pm\frac{A(2I+1)}{4}\sqrt{1+\frac{4m_{F}}{2I+1}x+x^{2}},
\end{split}
\end{equation}
where $x$$=$$\frac{2(g_J-g_I)\mu_B}{A(2I+1)} B$, $\mu_B$ is the Bohr-magneton, $+(-)$ refers to the energy levels in the ground-state hyperfine manifolds with $F$$=$$9/2(7/2)$, $g_J$ and $g_I$ are the Land\'{e} $g$ factor of the electron and the nucleus, respectively~\cite{Daniel2019K40}.
Using Eq.~(\ref{Breit}), we obtain a set of coupled equations for the two transitions:
\begin{eqnarray}\label{fangcheng}
\begin{aligned}
\omega_{\alpha}-\omega_{\beta}=&2\mu_{B}g_{I}B/\hbar,
 \\
\omega_{\alpha}+\omega_{\beta} =&-\frac{A(2I+1)}{2\hbar}\Bigg(\sqrt{1+\frac{2}{2I+1}x+x^{2}}\\
&+\sqrt{1-\frac{2}{2I+1}x+x^{2}}\Bigg),
\end{aligned}
\end{eqnarray}
where $\omega_{\alpha/\beta}$ denotes the transition angular frequency. According to Eq.~(\ref{fangcheng}), one can find that the value of the magnetic field can be replaced by the frequency difference between the two transitions. Furthermore, we obtain an magnetic-field-calibration-free expression for the hyperfine constant $A$ as
\begin{eqnarray}\label{freeconstant}
\begin{aligned}
A=-2\hbar \omega_{0} \sqrt{\frac{1-M^{2}}{(2I+1)^{2}-M^{2}}}
\end{aligned},
\end{eqnarray}
where $\omega_{0}=(\omega_{\alpha}+\omega_{\beta})/2$ and $M=\frac{g_J-g_I}{g_I}\frac{\omega_{\alpha}-\omega_{\beta}}{\omega_{\alpha}+\omega_{\beta}}$ (see Sec.~I of the Supplemental Material \cite{SM2026}).

\begin{figure}[t]
\includegraphics[width=3.3in]{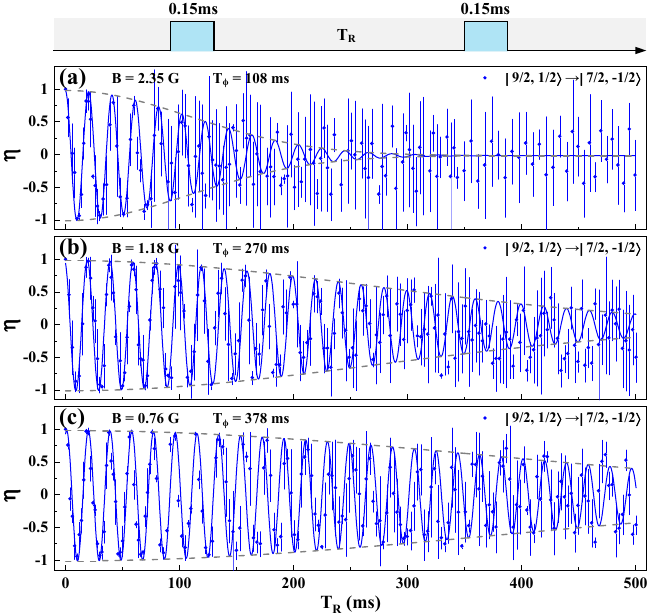}
\caption{Ramsey oscillations on the transition $\beta$ as a function of the free-evolution time $T_R$ at different bias magnetic fields, measured using a square $\pi/2$ pulse with duration $\tau_p=0.15$~ms for state preparation and readout.
The solid lines are fits using a Gaussian phase-diffusion model, while the dashed lines indicate the corresponding Ramsey contrast decay envelopes. Throughout this Letter, each data point represents the average of three independent measurements, and error bars denote one standard deviation.
\label{Ramsey}}
\end{figure}

A detailed description of the experimental setup is available in Refs.~\cite{WeiDong2007CPL,chaiShijie2012ActaSinQuantumOpt,Weidong2006ActaPhysSin,Wangaqiong2016ASQO,liangchao2017ASQO,MiaoJie2022CPB}.
Initially, approximately $N = 4\times10^{6}$ ultracold $^{40}$K atoms are prepared in the hyperfine Zeeman states $\lvert 9/2, -1/2 \rangle$ and $\lvert 9/2, 1/2 \rangle$ for the $\alpha$ and $\beta$ transitions, respectively, at a temperature of $0.3\,T_F$ in a crossed optical dipole trap (ODT), forming a deeply degenerate Fermi gas. Here the Fermi temperature is defined as $T_F = \hbar \bar{\omega} (6N)^{1/3} / k_B$, with $\bar{\omega}$ the geometric mean trapping frequency and $k_B$ the Boltzmann constant~\cite{Ding2024CPB}. Then the $\alpha$ and $\beta$ transitions transfer atoms from the $F=9/2$ manifold to $\lvert 7/2, 1/2 \rangle$ and $\lvert 7/2, -1/2 \rangle$, respectively. To fully determine the populations of the initial and final states, an adiabatic Landau–Zener transition is used to transfer atoms from the $F=7/2$ manifold back into a distinct Zeeman sublevel of the $F=9/2$ manifold. By using time-of-flight absorption imaging with a Stern–Gerlach gradient magnetic field, we can simultaneously resolve the populations originating from both the initial and final states of the transitions. Finally, the transition frequencies are determined by locating the resonance condition, identified by the maximum of the population imbalance $\eta = (n_{7/2} - n_{9/2})/(n_{7/2} + n_{9/2})$, where $n_F$ ($F = 7/2, 9/2$) denotes the number of atoms in the respective Zeeman sublevel.

\begin{figure*}[t]
\includegraphics[width=5.0in]{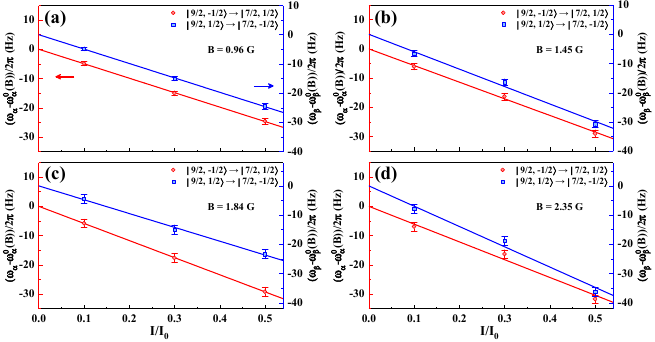}
\caption{Resonance frequencies of the $\alpha$ and $\beta$ transitions versus ODT intensity at different bias magnetic fields. The $\alpha$ ($\beta$) data are plotted against the left (right) vertical axis. Here, $\omega_{\alpha(\beta)}$ denotes the resonance frequency at the given magnetic field and ODT intensity, and $\omega_{\alpha(\beta)}^{0}$ is the corresponding resonance frequency at $I=0$ obtained by linear extrapolation. $I_0$ denotes the reference intensity at the center of the ODT formed by two mutually incoherent linearly polarized 1064~nm laser beams propagating in orthogonal directions, with powers of $3.5\,\mathrm{W}$ and $2.5\,\mathrm{W}$ and waists of approximately 45\,$\mu\mathrm{m}$. Different trap intensities are obtained by proportional scaling of the laser powers at fixed beam waist.
\label{shift}}
\end{figure*}

To measure the hyperfine constant $A$ with high precision, we first perform microwave spectroscopy using a Gaussian-shaped $\pi$ pulse to initially determine the two magnetically insensitive transition frequencies $\omega_{\alpha}$ and $\omega_{\beta}$~\cite{Shan2026PRA}, as shown in Fig.~\ref{setup}(b).
We then determine these frequencies with much higher precision by employing Ramsey oscillations using two square microwave pulses of 0.15~ms duration, separated by a free evolution time $T_R$ (see Sec. II of the Supplemental Material~\cite{SM2026}). Because the transitions are approximately first-order magnetically insensitive, the transition frequency uncertainty is dominated by residual higher-order Zeeman shifts induced by magnetic-field noise.
In our current experiment, with magnetic-field noise at approximately the 100 $\mu$G level, the Ramsey coherence time can reach several hundred milliseconds, as shown in Fig. 2, enabling Hertz-level frequency uncertainty.
Moreover, a striking enhancement of coherence is observed as the magnetic field is reduced. At even lower magnetic fields, however, the coherence is instead limited by the near-degeneracy of hyperfine Zeeman sublevels and by state mixing induced by spin-exchange collisions~\cite{Peng2018CPL}.
By fitting the Ramsey fringes with a Gaussian phase-diffusion Ramsey model, we determine the transition angular frequency and the coherence time $T_\phi$, where $T_\phi$ is defined as the Ramsey free-evolution time at which the rms accumulated phase uncertainty reaches one radian, i.e., $\sigma_{\phi}=\sigma_{\omega}T_\phi=1$.
The transition angular frequency uncertainty induced by magnetic field noise is then estimated as $\sigma_{\omega}=1/T_\phi$ (see Secs. III and V A of the Supplemental Material~\cite{SM2026}).

Although ODTs offer the advantage of producing and maintaining high-density atomic samples over the entire interrogation time compared with traditional fountain clock methods, they also induce an ac-Stark shift that affects the precise measurement of the transition frequencies~\cite{Porto2010PRA, Zhan2016PRL}. In our experiment, the ODT is formed by far-detuned linearly polarized laser beams. In the low-intensity range, the dominant light shift arises from the second-order AC Stark effect and is therefore expected to scale linearly with optical intensity~\cite{DSteck}. This allows us to extrapolate the resonance frequency to zero optical intensity via a linear fit, thereby effectively eliminating the trap-induced AC Stark frequency shift. In Fig.~\ref{shift}, we measure the light-induced shifts at different magnetic fields when the ODT intensity is varied, and extrapolate them to zero intensity by linear fitting. The extrapolation uncertainty is mainly dominated by the covariance of the fitting procedure and the statistical uncertainties from repeated measurements at each data point (see Secs.~IV A and V B of the Supplemental Material~\cite{SM2026}).

\begin{figure}[t]
\includegraphics[width=2.3in]{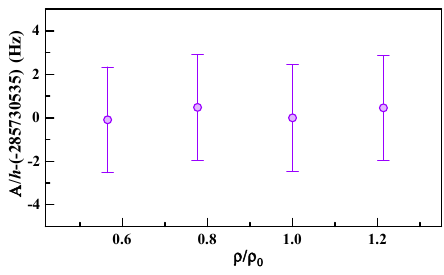}
\caption{Measured hyperfine constant $A$ at different atomic densities. The bias magnetic field is fixed at $B=1.06$~G and the ODT is maintained at $I=0.1I_0$. $\rho_0=8\times10^{13}\,\mathrm{cm}^{-3}$ denotes the reference density at the center of the ODT for $N=4\times10^6$, while different densities are achieved by varying the atom number.
\label{density}}
\end{figure}

The use of a trapped degenerate Fermi gas is not expected to introduce a significant interaction-induced modification in the measured hyperfine constant $A$. This is because s-wave collisions are strongly suppressed by the Pauli exclusion principle, while interaction-induced shifts of the transition frequencies are absent due to the invariance of fermionic contact interactions under coherent driving~\cite{Gupta2003Science,Zwierlein2003PRL}. Experimentally, this expectation is confirmed by the atomic density scan shown in Fig.~\ref{density}, which shows no systematic dependence of the extracted hyperfine constant $A$ on atomic density. The observed variations in $A$ are limited to random fluctuations at the sub-Hz level, well below the measurement uncertainty (see Sec.~V~C of the Supplemental Material~\cite{SM2026}). In addition, an accurate determination of the hyperfine constant \( A \) also depends on the precise experimental values of the Land\'{e} \( g \) factors \( g_J \) and \( g_I \). At our current level of experimental precision, the available values of \( g_J = 2.00229421(24) \) and \( g_I = 0.000176490(34) \)~\cite{Arimondo1977RMP, Daniel2019K40} are sufficient for an accurate determination of $A$ (see Sec. V D of the Supplemental Material~\cite{SM2026}).

\begin{figure}[t]
\includegraphics[width=3.2in]{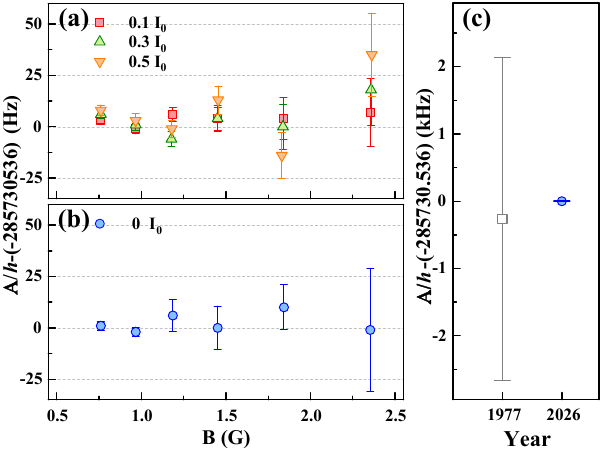}
\caption{(a)~Measured hyperfine constant $A$ at different magnetic fields and ODT intensities. (b)~Measured hyperfine constant $A$ after removing the ODT-induced ac Stark shift. (c)~Hyperfine constant determined in this work compared with previous measurements reported in Refs.~\cite{Zacharias1942PR, Davis1949PR, Eisinger1952PR, Arimondo1977RMP}.
The gray square represents the previously recommended value $-h\times285.7308(24)\,\mathrm{MHz}$ in 1977~\cite{Arimondo1977RMP}, while the blue dot shows the present value $A = -h\times 285.730536(2)\,\mathrm{MHz}$.
\label{constant}}
\end{figure}

Finally, based on the measured frequencies of the two magnetically insensitive transitions, the hyperfine constant $A$ can be determined according to Eq.~(\ref{freeconstant}), as shown in Fig.~\ref{constant}. It is found that the hyperfine constant $A$ extracted at lower trap intensities exhibits greater stability against variations in the magnetic field, as illustrated in Fig.~\ref{constant}(a). Furthermore, a more precise value of $A$ can be obtained by eliminating the trap-induced ac Stark shift via zero-intensity extrapolation, as shown in Fig.~\ref{constant}(b). Using the data in Fig.~\ref{constant}(b) and an inverse-variance weighted mean, the hyperfine structure constant of $^{40}$K is determined to be $A$$=$$ -h\times 285.730536(2)\,\text{MHz}$. The 2~Hz uncertainty is obtained by accounting for contributions from magnetic-field noise, AC Stark shifts, and uncertainties in the Land\'{e} \( g \) factors, as well as the negligible interaction-induced shifts. Additional details on the precise determination and uncertainty budget of hyperfine constant $A$ are provided in Secs.~IV and V of the Supplemental Material~\cite{SM2026}. This value represents an improvement in precision of approximately three orders of magnitude over previously reported values~\cite{Allegrini2022JPC}, as presented in Fig.~\ref{constant}(c).

In conclusion, we develop a magnetic-field-calibration-free method to determine the hyperfine constant. The key idea is the simultaneous measurement of two specifically chosen magnetically insensitive transitions, from which the hyperfine constant can be directly extracted using only the measured transition frequencies. The magnetic-field insensitivity ensures Hertz-level precision in the frequency measurements. This enables the experiment to be performed without the magnetic shielding and magnetic-field calibration required in conventional methods. In contrast to conventional fountain-clock and atomic-beam approaches, we prepare a high-density ultracold atomic sample in the degenerate regime confined in an optical dipole trap, which significantly enhances the atomic density.
Under these experimental conditions, we employ Ramsey interferometry to measure the two transition frequencies, leading to a determination of the hyperfine structure constant, \(A =-h\times 285.730536(2)\text{ MHz}\). This result improves the most precise previously reported value by three orders of magnitude~\cite{Zacharias1942PR, Davis1949PR, Eisinger1952PR, Arimondo1977RMP}.

Our current work brings the measurement of the hyperfine constant of \({}^{40}\mathrm{K}\) on the Hertz-level precision footing of other alkali-metal isotopes, thereby resolving a long-standing problem of low measurement precision for \({}^{40}\mathrm{K}\) that has persisted for more than seventy years. The refined value of the hyperfine constant provides a more reliable atomic reference for precision spectroscopy, which can significantly improve applications in magnetic-field calibration~\cite{Rabi1931PR, Regal2004PRL}, Feshbach-resonance fitting~\cite{Tiesinga2010RMP, Wille2008PRL, Zwierlein2012PRA}, and the determination of spin-dependent interaction strengths~\cite{Jin2003PRL}. Moreover, this approach can be naturally extended to other atoms and hyperfine transitions that are important for quantum precision measurement and quantum sensing.

It should be noted that the two magnetically insensitive transitions employed in our scheme only eliminate the first-order differential Zeeman shift.
The transition frequencies are still affected by higher-order Zeeman effects, which limit the atomic coherence time in Ramsey interferometry and make the measurements susceptible to magnetic field noise, thereby confining the precision to the Hertz level. Further improvements, such as enhanced magnetic shielding and increased magnetic-field stability, could prolong the coherence time and enable the measurement precision of the hyperfine constant to reach even higher levels.
In the current experiment, the ODT-induced AC Stark shift is carefully characterized and compensated to minimize its impact on the measurements. In principle, the AC Stark shift is determined by the scalar, vector, and tensor polarizabilities of the atoms~\cite{Porto2010PRA, Mitroy2013PRA, Zhan2016PRL, Widera2016PRA, Tiancai2019PRL, Wen2021JOSAB}, which depend on the wavelength and polarization of the laser.
In future experiments, by choosing an appropriate laser wavelength and polarization, it may be possible to eliminate the differential
light shift of the clock transitions fundamentally.

\begin{acknowledgments}
This research is supported by National Key Research and Development Program of China (Grants No.~2022YFA1404101, and No.~2021YFA1401700), Quantum Science and Technology-National Science and Technology Major Project (Grants  No.~2021ZD0302003), National Natural Science Foundation of China (Grants No.~12034011, No.~92476001, No.~U23A6004, {No.~12488301}, No.~12474266, No.~12474252, No.~12374245, No.~12322409, and No.~92576205, and No.~12504336), the Fundamental Research Program of Shanxi Province (No.~202403021221001), and the Fund Program for the Scientific Activities of Selected Returned Overseas Professionals in Shanxi Province(No.~20250004).
\end{acknowledgments}

\bibliography{reference-Hyperfine-Constant}

\begin{thebibliography}{49}%
\makeatletter
\providecommand \@ifxundefined [1]{%
 \@ifx{#1\undefined}
}%
\providecommand \@ifnum [1]{%
 \ifnum #1\expandafter \@firstoftwo
 \else \expandafter \@secondoftwo
 \fi
}%
\providecommand \@ifx [1]{%
 \ifx #1\expandafter \@firstoftwo
 \else \expandafter \@secondoftwo
 \fi
}%
\providecommand \natexlab [1]{#1}%
\providecommand \enquote  [1]{``#1''}%
\providecommand \bibnamefont  [1]{#1}%
\providecommand \bibfnamefont [1]{#1}%
\providecommand \citenamefont [1]{#1}%
\providecommand \href@noop [0]{\@secondoftwo}%
\providecommand \href [0]{\begingroup \@sanitize@url \@href}%
\providecommand \@href[1]{\@@startlink{#1}\@@href}%
\providecommand \@@href[1]{\endgroup#1\@@endlink}%
\providecommand \@sanitize@url [0]{\catcode `\\12\catcode `\$12\catcode
  `\&12\catcode `\#12\catcode `\^12\catcode `\_12\catcode `\%12\relax}%
\providecommand \@@startlink[1]{}%
\providecommand \@@endlink[0]{}%
\providecommand \url  [0]{\begingroup\@sanitize@url \@url }%
\providecommand \@url [1]{\endgroup\@href {#1}{\urlprefix }}%
\providecommand \urlprefix  [0]{URL }%
\providecommand \Eprint [0]{\href }%
\providecommand \doibase [0]{https://doi.org/}%
\providecommand \selectlanguage [0]{\@gobble}%
\providecommand \bibinfo  [0]{\@secondoftwo}%
\providecommand \bibfield  [0]{\@secondoftwo}%
\providecommand \translation [1]{[#1]}%
\providecommand \BibitemOpen [0]{}%
\providecommand \bibitemStop [0]{}%
\providecommand \bibitemNoStop [0]{.\EOS\space}%
\providecommand \EOS [0]{\spacefactor3000\relax}%
\providecommand \BibitemShut  [1]{\csname bibitem#1\endcsname}%
\let\auto@bib@innerbib\@empty
%</preamble>
\bibitem [{\citenamefont {Ramsey}(1956)}]{Ramsey1956Beams}%
  \BibitemOpen
  \bibfield  {author} {\bibinfo {author} {\bibfnamefont {N.~F.}\ \bibnamefont
  {Ramsey}},\ }\href@noop {} {\emph {\bibinfo {title} {Molecular Beams}}}\
  (\bibinfo  {publisher} {Oxford University Press},\ \bibinfo {address} {New
  York},\ \bibinfo {year} {1956})\BibitemShut {NoStop}%
\bibitem [{\citenamefont {Metcalf}\ and\ \citenamefont {van~der
  Straten}(1999)}]{Metcalf1999Fountain}%
  \BibitemOpen
  \bibfield  {author} {\bibinfo {author} {\bibfnamefont {H.~J.}\ \bibnamefont
  {Metcalf}}\ and\ \bibinfo {author} {\bibfnamefont {P.}~\bibnamefont {van~der
  Straten}},\ }\href@noop {} {\emph {\bibinfo {title} {Laser Cooling and
  Trapping of Neutral Atoms}}}\ (\bibinfo  {publisher} {Springer-Verlag},\
  \bibinfo {address} {New York},\ \bibinfo {year} {1999})\BibitemShut {NoStop}%
\bibitem [{\citenamefont {Udem}\ \emph {et~al.}(2002)\citenamefont {Udem},
  \citenamefont {Holzwarth},\ and\ \citenamefont
  {H{\"a}nsch}}]{Hansch2002Nature}%
  \BibitemOpen
  \bibfield  {author} {\bibinfo {author} {\bibfnamefont {T.}~\bibnamefont
  {Udem}}, \bibinfo {author} {\bibfnamefont {R.}~\bibnamefont {Holzwarth}},\
  and\ \bibinfo {author} {\bibfnamefont {T.~W.}\ \bibnamefont {H{\"a}nsch}},\
  }\bibfield  {title} {\bibinfo {title} {Optical frequency metrology},\ }\href
  {https://doi.org/10.1038/416233a} {\bibfield  {journal} {\bibinfo  {journal}
  {Nature}\ }\textbf {\bibinfo {volume} {416}},\ \bibinfo {pages} {233}
  (\bibinfo {year} {2002})}\BibitemShut {NoStop}%
\bibitem [{\citenamefont {Schlecht}\ and\ \citenamefont
  {McColm}(1966)}]{Schlecht1966LiHFS}%
  \BibitemOpen
  \bibfield  {author} {\bibinfo {author} {\bibfnamefont {R.~G.}\ \bibnamefont
  {Schlecht}}\ and\ \bibinfo {author} {\bibfnamefont {D.~W.}\ \bibnamefont
  {McColm}},\ }\bibfield  {title} {\bibinfo {title} {{Hyperfine Structure of
  the Stable Lithium Isotopes. I}},\ }\href
  {https://doi.org/10.1103/PhysRev.142.11} {\bibfield  {journal} {\bibinfo
  {journal} {Phys. Rev.}\ }\textbf {\bibinfo {volume} {142}},\ \bibinfo {pages}
  {11} (\bibinfo {year} {1966})}\BibitemShut {NoStop}%
\bibitem [{\citenamefont {Arditi}\ and\ \citenamefont
  {Carver}(1958)}]{Arditi1958NaHFS}%
  \BibitemOpen
  \bibfield  {author} {\bibinfo {author} {\bibfnamefont {M.}~\bibnamefont
  {Arditi}}\ and\ \bibinfo {author} {\bibfnamefont {T.~R.}\ \bibnamefont
  {Carver}},\ }\bibfield  {title} {\bibinfo {title} {{Optical Detection of
  Zero-Field Hyperfine Splitting of ${}^{23}\mathrm{Na}$}},\ }\href
  {https://doi.org/10.1103/PhysRev.109.1012} {\bibfield  {journal} {\bibinfo
  {journal} {Phys. Rev.}\ }\textbf {\bibinfo {volume} {109}},\ \bibinfo {pages}
  {1012} (\bibinfo {year} {1958})}\BibitemShut {NoStop}%
\bibitem [{\citenamefont {Beckmann}\ \emph {et~al.}(1974)\citenamefont
  {Beckmann}, \citenamefont {B{\"o}klen},\ and\ \citenamefont
  {Elke}}]{beckmann1974ZFP}%
  \BibitemOpen
  \bibfield  {author} {\bibinfo {author} {\bibfnamefont {A.}~\bibnamefont
  {Beckmann}}, \bibinfo {author} {\bibfnamefont {K.~D.}\ \bibnamefont
  {B{\"o}klen}},\ and\ \bibinfo {author} {\bibfnamefont {D.}~\bibnamefont
  {Elke}},\ }\bibfield  {title} {\bibinfo {title} {{Precision measurements of
  the nuclear magnetic dipole moments of $^{6}$Li, $^{7}$Li, $^{23}$Na,
  $^{39}$K and $^{41}$K}},\ }\href {https://doi.org/10.1007/BF01680407}
  {\bibfield  {journal} {\bibinfo  {journal} {Zeitschrift f{\"u}r Physik}\
  }\textbf {\bibinfo {volume} {270}},\ \bibinfo {pages} {173} (\bibinfo {year}
  {1974})}\BibitemShut {NoStop}%
\bibitem [{\citenamefont {Peper}\ \emph {et~al.}(2019)\citenamefont {Peper},
  \citenamefont {Helmrich}, \citenamefont {Butscher}, \citenamefont {Agner},
  \citenamefont {Schmutz}, \citenamefont {Merkt},\ and\ \citenamefont
  {Deiglmayr}}]{Peper2019K39GroundHFS}%
  \BibitemOpen
  \bibfield  {author} {\bibinfo {author} {\bibfnamefont {M.}~\bibnamefont
  {Peper}}, \bibinfo {author} {\bibfnamefont {F.}~\bibnamefont {Helmrich}},
  \bibinfo {author} {\bibfnamefont {J.}~\bibnamefont {Butscher}}, \bibinfo
  {author} {\bibfnamefont {J.~A.}\ \bibnamefont {Agner}}, \bibinfo {author}
  {\bibfnamefont {H.}~\bibnamefont {Schmutz}}, \bibinfo {author} {\bibfnamefont
  {F.}~\bibnamefont {Merkt}},\ and\ \bibinfo {author} {\bibfnamefont
  {J.}~\bibnamefont {Deiglmayr}},\ }\bibfield  {title} {\bibinfo {title}
  {Precision measurement of the ionization energy and quantum defects of
  ${}^{39}\mathrm{K\,I}$},\ }\href
  {https://doi.org/10.1103/PhysRevA.100.012501} {\bibfield  {journal} {\bibinfo
   {journal} {Phys. Rev. A}\ }\textbf {\bibinfo {volume} {100}},\ \bibinfo
  {pages} {012501} (\bibinfo {year} {2019})}\BibitemShut {NoStop}%
\bibitem [{\citenamefont {Bender}\ \emph {et~al.}(1958)\citenamefont {Bender},
  \citenamefont {Beaty},\ and\ \citenamefont {Chi}}]{Bender1958Rb87HFS}%
  \BibitemOpen
  \bibfield  {author} {\bibinfo {author} {\bibfnamefont {P.~L.}\ \bibnamefont
  {Bender}}, \bibinfo {author} {\bibfnamefont {E.~C.}\ \bibnamefont {Beaty}},\
  and\ \bibinfo {author} {\bibfnamefont {A.~R.}\ \bibnamefont {Chi}},\
  }\bibfield  {title} {\bibinfo {title} {Optical detection of narrow
  ${}^{87}\mathrm{Rb}$ hyperfine absorption lines},\ }\href
  {https://doi.org/10.1103/PhysRevLett.1.311} {\bibfield  {journal} {\bibinfo
  {journal} {Phys. Rev. Lett.}\ }\textbf {\bibinfo {volume} {1}},\ \bibinfo
  {pages} {311} (\bibinfo {year} {1958})}\BibitemShut {NoStop}%
\bibitem [{\citenamefont {Wang}\ \emph {et~al.}(2019)\citenamefont {Wang},
  \citenamefont {Zhang}, \citenamefont {Guang}, \citenamefont {Zhang},
  \citenamefont {Wang}, \citenamefont {Wei},\ and\ \citenamefont
  {Wang}}]{Yuzhu2019PRA}%
  \BibitemOpen
  \bibfield  {author} {\bibinfo {author} {\bibfnamefont {Q.}~\bibnamefont
  {Wang}}, \bibinfo {author} {\bibfnamefont {N.}~\bibnamefont {Zhang}},
  \bibinfo {author} {\bibfnamefont {W.}~\bibnamefont {Guang}}, \bibinfo
  {author} {\bibfnamefont {S.}~\bibnamefont {Zhang}}, \bibinfo {author}
  {\bibfnamefont {W.}~\bibnamefont {Wang}}, \bibinfo {author} {\bibfnamefont
  {R.}~\bibnamefont {Wei}},\ and\ \bibinfo {author} {\bibfnamefont
  {Y.}~\bibnamefont {Wang}},\ }\bibfield  {title} {\bibinfo {title} {Precision
  measurements of the ground-state hyperfine splitting of $^{85}\mathrm{Rb}$
  using an atomic fountain clock},\ }\href
  {https://doi.org/10.1103/PhysRevA.100.022510} {\bibfield  {journal} {\bibinfo
   {journal} {Phys. Rev. A}\ }\textbf {\bibinfo {volume} {100}},\ \bibinfo
  {pages} {022510} (\bibinfo {year} {2019})}\BibitemShut {NoStop}%
\bibitem [{\citenamefont {Ovchinnikov}\ \emph {et~al.}(2015)\citenamefont
  {Ovchinnikov}, \citenamefont {Szymaniec},\ and\ \citenamefont
  {Edris}}]{ovchinnikov2015Metrologia}%
  \BibitemOpen
  \bibfield  {author} {\bibinfo {author} {\bibfnamefont {Y.~B.}\ \bibnamefont
  {Ovchinnikov}}, \bibinfo {author} {\bibfnamefont {K.}~\bibnamefont
  {Szymaniec}},\ and\ \bibinfo {author} {\bibfnamefont {S.}~\bibnamefont
  {Edris}},\ }\bibfield  {title} {\bibinfo {title} {{Measurement of Rubidium
  Ground-State Hyperfine Transition Frequency Using Atomic Fountains}},\ }\href
  {https://doi.org/10.1088/0026-1394/52/4/595} {\bibfield  {journal} {\bibinfo
  {journal} {Metrologia}\ }\textbf {\bibinfo {volume} {52}},\ \bibinfo {pages}
  {595} (\bibinfo {year} {2015})}\BibitemShut {NoStop}%
\bibitem [{\citenamefont {Essen}\ and\ \citenamefont
  {Parry}(1955)}]{Essen1955CsClock}%
  \BibitemOpen
  \bibfield  {author} {\bibinfo {author} {\bibfnamefont {L.}~\bibnamefont
  {Essen}}\ and\ \bibinfo {author} {\bibfnamefont {J.~V.~L.}\ \bibnamefont
  {Parry}},\ }\bibfield  {title} {\bibinfo {title} {An atomic standard of
  frequency and time interval: A caesium resonator},\ }\href
  {https://doi.org/10.1038/176280a0} {\bibfield  {journal} {\bibinfo  {journal}
  {Nature}\ }\textbf {\bibinfo {volume} {176}},\ \bibinfo {pages} {280}
  (\bibinfo {year} {1955})}\BibitemShut {NoStop}%
\bibitem [{\citenamefont {Markowitz}\ \emph {et~al.}(1958)\citenamefont
  {Markowitz}, \citenamefont {Hall}, \citenamefont {Essen},\ and\ \citenamefont
  {Parry}}]{Parry1958PRL}%
  \BibitemOpen
  \bibfield  {author} {\bibinfo {author} {\bibfnamefont {W.}~\bibnamefont
  {Markowitz}}, \bibinfo {author} {\bibfnamefont {R.~G.}\ \bibnamefont {Hall}},
  \bibinfo {author} {\bibfnamefont {L.}~\bibnamefont {Essen}},\ and\ \bibinfo
  {author} {\bibfnamefont {J.~V.~L.}\ \bibnamefont {Parry}},\ }\bibfield
  {title} {\bibinfo {title} {Frequency of cesium in terms of ephemeris time},\
  }\href {https://doi.org/10.1103/PhysRevLett.1.105} {\bibfield  {journal}
  {\bibinfo  {journal} {Phys. Rev. Lett.}\ }\textbf {\bibinfo {volume} {1}},\
  \bibinfo {pages} {105} (\bibinfo {year} {1958})}\BibitemShut {NoStop}%
\bibitem [{\citenamefont {Arimondo}\ \emph {et~al.}(1977)\citenamefont
  {Arimondo}, \citenamefont {Inguscio},\ and\ \citenamefont
  {Violino}}]{Arimondo1977RMP}%
  \BibitemOpen
  \bibfield  {author} {\bibinfo {author} {\bibfnamefont {E.}~\bibnamefont
  {Arimondo}}, \bibinfo {author} {\bibfnamefont {M.}~\bibnamefont {Inguscio}},\
  and\ \bibinfo {author} {\bibfnamefont {P.}~\bibnamefont {Violino}},\
  }\bibfield  {title} {\bibinfo {title} {Experimental determinations of the
  hyperfine structure in the alkali atoms},\ }\href
  {https://doi.org/10.1103/RevModPhys.49.31} {\bibfield  {journal} {\bibinfo
  {journal} {Rev. Mod. Phys.}\ }\textbf {\bibinfo {volume} {49}},\ \bibinfo
  {pages} {31} (\bibinfo {year} {1977})}\BibitemShut {NoStop}%
\bibitem [{\citenamefont {Allegrini}\ \emph {et~al.}(2022)\citenamefont
  {Allegrini}, \citenamefont {Arimondo},\ and\ \citenamefont
  {Orozco}}]{Allegrini2022JPC}%
  \BibitemOpen
  \bibfield  {author} {\bibinfo {author} {\bibfnamefont {M.}~\bibnamefont
  {Allegrini}}, \bibinfo {author} {\bibfnamefont {E.}~\bibnamefont
  {Arimondo}},\ and\ \bibinfo {author} {\bibfnamefont {L.~A.}\ \bibnamefont
  {Orozco}},\ }\bibfield  {title} {\bibinfo {title} {Survey of hyperfine
  structure measurements in alkali atoms},\ }\href
  {https://doi.org/10.1063/5.0098061} {\bibfield  {journal} {\bibinfo
  {journal} {J. Phys. Chem. Ref. Data}\ }\textbf {\bibinfo {volume} {51}},\
  \bibinfo {pages} {043102} (\bibinfo {year} {2022})}\BibitemShut {NoStop}%
\bibitem [{\citenamefont {Wei}\ \emph {et~al.}(2007{\natexlab{a}})\citenamefont
  {Wei}, \citenamefont {Xiong}, \citenamefont {Chen},\ and\ \citenamefont
  {Zhang}}]{Wei2007Enriched40K}%
  \BibitemOpen
  \bibfield  {author} {\bibinfo {author} {\bibfnamefont {D.}~\bibnamefont
  {Wei}}, \bibinfo {author} {\bibfnamefont {D.-Z.}\ \bibnamefont {Xiong}},
  \bibinfo {author} {\bibfnamefont {H.-X.}\ \bibnamefont {Chen}},\ and\
  \bibinfo {author} {\bibfnamefont {J.}~\bibnamefont {Zhang}},\ }\bibfield
  {title} {\bibinfo {title} {{An Enriched {$^{40}$}K Source for Atomic
  Cooling}},\ }\href {https://doi.org/10.1088/0256-307X/24/3/025} {\bibfield
  {journal} {\bibinfo  {journal} {Chin. Phys. Lett.}\ }\textbf {\bibinfo
  {volume} {24}},\ \bibinfo {pages} {679} (\bibinfo {year}
  {2007}{\natexlab{a}})}\BibitemShut {NoStop}%
\bibitem [{\citenamefont {Meija}\ \emph {et~al.}(2016)\citenamefont {Meija},
  \citenamefont {Coplen}, \citenamefont {Berglund}, \citenamefont {Brand},
  \citenamefont {De~Bi{\`e}vre}, \citenamefont {Gr{\"o}ning}, \citenamefont
  {Holden}, \citenamefont {Irrgeher}, \citenamefont {Loss}, \citenamefont
  {Walczyk},\ and\ \citenamefont {Prohaska}}]{Prohaska2016PAC}%
  \BibitemOpen
  \bibfield  {author} {\bibinfo {author} {\bibfnamefont {J.}~\bibnamefont
  {Meija}}, \bibinfo {author} {\bibfnamefont {T.~B.}\ \bibnamefont {Coplen}},
  \bibinfo {author} {\bibfnamefont {M.}~\bibnamefont {Berglund}}, \bibinfo
  {author} {\bibfnamefont {W.~A.}\ \bibnamefont {Brand}}, \bibinfo {author}
  {\bibfnamefont {P.}~\bibnamefont {De~Bi{\`e}vre}}, \bibinfo {author}
  {\bibfnamefont {M.}~\bibnamefont {Gr{\"o}ning}}, \bibinfo {author}
  {\bibfnamefont {N.~E.}\ \bibnamefont {Holden}}, \bibinfo {author}
  {\bibfnamefont {J.}~\bibnamefont {Irrgeher}}, \bibinfo {author}
  {\bibfnamefont {R.~D.}\ \bibnamefont {Loss}}, \bibinfo {author}
  {\bibfnamefont {T.}~\bibnamefont {Walczyk}},\ and\ \bibinfo {author}
  {\bibfnamefont {T.}~\bibnamefont {Prohaska}},\ }\bibfield  {title} {\bibinfo
  {title} {Isotopic compositions of the elements 2013 ({IUPAC} technical
  report)},\ }\href {https://doi.org/10.1515/pac-2015-0503} {\bibfield
  {journal} {\bibinfo  {journal} {Pure Appl. Chem.}\ }\textbf {\bibinfo
  {volume} {88}},\ \bibinfo {pages} {293} (\bibinfo {year} {2016})}\BibitemShut
  {NoStop}%
\bibitem [{\citenamefont {Zacharias}(1942)}]{Zacharias1942PR}%
  \BibitemOpen
  \bibfield  {author} {\bibinfo {author} {\bibfnamefont {J.~R.}\ \bibnamefont
  {Zacharias}},\ }\bibfield  {title} {\bibinfo {title} {The {{Nuclear Spin}}
  and {{Magnetic Moment}} of {{K}}$^{40}$},\ }\href
  {https://doi.org/10.1103/PhysRev.61.270} {\bibfield  {journal} {\bibinfo
  {journal} {Phys. Rev.}\ }\textbf {\bibinfo {volume} {61}},\ \bibinfo {pages}
  {270} (\bibinfo {year} {1942})}\BibitemShut {NoStop}%
\bibitem [{\citenamefont {Davis}\ \emph {et~al.}(1949)\citenamefont {Davis},
  \citenamefont {Nagle},\ and\ \citenamefont {Zacharias}}]{Davis1949PR}%
  \BibitemOpen
  \bibfield  {author} {\bibinfo {author} {\bibfnamefont {L.}~\bibnamefont
  {Davis}}, \bibinfo {author} {\bibfnamefont {D.~E.}\ \bibnamefont {Nagle}},\
  and\ \bibinfo {author} {\bibfnamefont {J.~R.}\ \bibnamefont {Zacharias}},\
  }\bibfield  {title} {\bibinfo {title} {Atomic {{Beam Magnetic Resonance
  Experiments}} with {{Radioactive Elements Na}}$^{22}$, {{K}}$^{40}$ ,
  {{Cs}}$^{135}$ , and {{Cs}}$^{137}$},\ }\href
  {https://doi.org/10.1103/PhysRev.76.1068} {\bibfield  {journal} {\bibinfo
  {journal} {Phys. Rev.}\ }\textbf {\bibinfo {volume} {76}},\ \bibinfo {pages}
  {1068} (\bibinfo {year} {1949})}\BibitemShut {NoStop}%
\bibitem [{\citenamefont {Eisinger}\ \emph {et~al.}(1952)\citenamefont
  {Eisinger}, \citenamefont {Bederson},\ and\ \citenamefont
  {Feld}}]{Eisinger1952PR}%
  \BibitemOpen
  \bibfield  {author} {\bibinfo {author} {\bibfnamefont {J.~T.}\ \bibnamefont
  {Eisinger}}, \bibinfo {author} {\bibfnamefont {B.}~\bibnamefont {Bederson}},\
  and\ \bibinfo {author} {\bibfnamefont {B.~T.}\ \bibnamefont {Feld}},\
  }\bibfield  {title} {\bibinfo {title} {The {{Magnetic Moment}} of
  {{K}}$^{40}$ and the {{Hyperfine Structure Anomaly}} of the {{Potassium
  Isotopes}}},\ }\href {https://doi.org/10.1103/PhysRev.86.73} {\bibfield
  {journal} {\bibinfo  {journal} {Phys. Rev.}\ }\textbf {\bibinfo {volume}
  {86}},\ \bibinfo {pages} {73} (\bibinfo {year} {1952})}\BibitemShut {NoStop}%
\bibitem [{\citenamefont {Falke}\ \emph {et~al.}(2006)\citenamefont {Falke},
  \citenamefont {Tiemann}, \citenamefont {Lisdat}, \citenamefont {Schnatz},\
  and\ \citenamefont {Grosche}}]{Falke2006PRA}%
  \BibitemOpen
  \bibfield  {author} {\bibinfo {author} {\bibfnamefont {S.}~\bibnamefont
  {Falke}}, \bibinfo {author} {\bibfnamefont {E.}~\bibnamefont {Tiemann}},
  \bibinfo {author} {\bibfnamefont {C.}~\bibnamefont {Lisdat}}, \bibinfo
  {author} {\bibfnamefont {H.}~\bibnamefont {Schnatz}},\ and\ \bibinfo {author}
  {\bibfnamefont {G.}~\bibnamefont {Grosche}},\ }\bibfield  {title} {\bibinfo
  {title} {{Transition frequencies of the D lines of $^{39}$K, $^{40}$K, and
  $^{41}$K measured with a femtosecond laser frequency comb}},\ }\href
  {https://doi.org/10.1103/PhysRevA.74.032503} {\bibfield  {journal} {\bibinfo
  {journal} {Phys. Rev. A}\ }\textbf {\bibinfo {volume} {74}},\ \bibinfo
  {pages} {032503} (\bibinfo {year} {2006})}\BibitemShut {NoStop}%
\bibitem [{\citenamefont {Behrle}\ \emph {et~al.}(2011)\citenamefont {Behrle},
  \citenamefont {Koschorreck},\ and\ \citenamefont {K{\"o}hl}}]{Behrle2011PRA}%
  \BibitemOpen
  \bibfield  {author} {\bibinfo {author} {\bibfnamefont {A.}~\bibnamefont
  {Behrle}}, \bibinfo {author} {\bibfnamefont {M.}~\bibnamefont
  {Koschorreck}},\ and\ \bibinfo {author} {\bibfnamefont {M.}~\bibnamefont
  {K{\"o}hl}},\ }\bibfield  {title} {\bibinfo {title} {Isotope shift and
  hyperfine splitting of the $4s$$\rightarrow$$5p$ transition in potassium},\
  }\href {https://doi.org/10.1103/PhysRevA.83.052507} {\bibfield  {journal}
  {\bibinfo  {journal} {Phys. Rev. A}\ }\textbf {\bibinfo {volume} {83}},\
  \bibinfo {pages} {052507} (\bibinfo {year} {2011})}\BibitemShut {NoStop}%
\bibitem [{\citenamefont {Hanley}\ \emph {et~al.}(2015)\citenamefont {Hanley},
  \citenamefont {Gregory}, \citenamefont {Hughes},\ and\ \citenamefont
  {Cornish}}]{Hanley2015JOSAB}%
  \BibitemOpen
  \bibfield  {author} {\bibinfo {author} {\bibfnamefont {R.~K.}\ \bibnamefont
  {Hanley}}, \bibinfo {author} {\bibfnamefont {P.~D.}\ \bibnamefont {Gregory}},
  \bibinfo {author} {\bibfnamefont {I.~G.}\ \bibnamefont {Hughes}},\ and\
  \bibinfo {author} {\bibfnamefont {S.~L.}\ \bibnamefont {Cornish}},\
  }\bibfield  {title} {\bibinfo {title} {{Absolute absorption on the potassium
  D lines: theory and experiment}},\ }\href
  {https://doi.org/10.1088/0953-4075/48/19/195004} {\bibfield  {journal}
  {\bibinfo  {journal} {J. Opt. Soc. Am. B}\ }\textbf {\bibinfo {volume}
  {48}},\ \bibinfo {pages} {195004} (\bibinfo {year} {2015})}\BibitemShut
  {NoStop}%
\bibitem [{\citenamefont {Wille}\ \emph
  {et~al.}(2008{\natexlab{a}})\citenamefont {Wille}, \citenamefont
  {Spiegelhalder}, \citenamefont {Kerner}, \citenamefont {Naik}, \citenamefont
  {Trenkwalder}, \citenamefont {Hendl}, \citenamefont {Schreck}, \citenamefont
  {Grimm}, \citenamefont {Tiecke}, \citenamefont {Walraven}, \citenamefont
  {Kokkelmans}, \citenamefont {Tiesinga},\ and\ \citenamefont
  {Julienne}}]{Julienne2008PRL}%
  \BibitemOpen
  \bibfield  {author} {\bibinfo {author} {\bibfnamefont {E.}~\bibnamefont
  {Wille}}, \bibinfo {author} {\bibfnamefont {F.~M.}\ \bibnamefont
  {Spiegelhalder}}, \bibinfo {author} {\bibfnamefont {G.}~\bibnamefont
  {Kerner}}, \bibinfo {author} {\bibfnamefont {D.}~\bibnamefont {Naik}},
  \bibinfo {author} {\bibfnamefont {A.}~\bibnamefont {Trenkwalder}}, \bibinfo
  {author} {\bibfnamefont {G.}~\bibnamefont {Hendl}}, \bibinfo {author}
  {\bibfnamefont {F.}~\bibnamefont {Schreck}}, \bibinfo {author} {\bibfnamefont
  {R.}~\bibnamefont {Grimm}}, \bibinfo {author} {\bibfnamefont {T.~G.}\
  \bibnamefont {Tiecke}}, \bibinfo {author} {\bibfnamefont {J.~T.~M.}\
  \bibnamefont {Walraven}}, \bibinfo {author} {\bibfnamefont {S.~J. J. M.~F.}\
  \bibnamefont {Kokkelmans}}, \bibinfo {author} {\bibfnamefont
  {E.}~\bibnamefont {Tiesinga}},\ and\ \bibinfo {author} {\bibfnamefont
  {P.~S.}\ \bibnamefont {Julienne}},\ }\bibfield  {title} {\bibinfo {title}
  {{Exploring an ultracold Fermi-Fermi mixture: Interspecies Feshbach
  resonances and scattering properties of $^{6}$Li and $^{40}$K}},\ }\href
  {https://doi.org/10.1103/PhysRevLett.100.053201} {\bibfield  {journal}
  {\bibinfo  {journal} {Phys. Rev. Lett.}\ }\textbf {\bibinfo {volume} {100}},\
  \bibinfo {pages} {053201} (\bibinfo {year} {2008}{\natexlab{a}})}\BibitemShut
  {NoStop}%
\bibitem [{\citenamefont {Park}\ \emph {et~al.}(2012)\citenamefont {Park},
  \citenamefont {Wu}, \citenamefont {Santiago}, \citenamefont {Tiecke},
  \citenamefont {Ahmadi},\ and\ \citenamefont {Zwierlein}}]{Zwierlein2012PRA}%
  \BibitemOpen
  \bibfield  {author} {\bibinfo {author} {\bibfnamefont {J.~W.}\ \bibnamefont
  {Park}}, \bibinfo {author} {\bibfnamefont {C.-H.}\ \bibnamefont {Wu}},
  \bibinfo {author} {\bibfnamefont {I.}~\bibnamefont {Santiago}}, \bibinfo
  {author} {\bibfnamefont {T.~G.}\ \bibnamefont {Tiecke}}, \bibinfo {author}
  {\bibfnamefont {P.}~\bibnamefont {Ahmadi}},\ and\ \bibinfo {author}
  {\bibfnamefont {M.~W.}\ \bibnamefont {Zwierlein}},\ }\bibfield  {title}
  {\bibinfo {title} {{Quantum degenerate Bose-Fermi mixture of chemically
  different atomic species with widely tunable interactions}},\ }\href
  {https://doi.org/10.1103/PhysRevA.85.051602} {\bibfield  {journal} {\bibinfo
  {journal} {Phys. Rev. A}\ }\textbf {\bibinfo {volume} {85}},\ \bibinfo
  {pages} {051602(R)} (\bibinfo {year} {2012})}\BibitemShut {NoStop}%
\bibitem [{\citenamefont {Shan}\ \emph {et~al.}(2026)\citenamefont {Shan},
  \citenamefont {Huang}, \citenamefont {Yang}, \citenamefont {Zhao},
  \citenamefont {Shen}, \citenamefont {Ye}, \citenamefont {Chen}, \citenamefont
  {Meng}, \citenamefont {Wang}, \citenamefont {Han},\ and\ \citenamefont
  {Zhang}}]{Shan2026PRA}%
  \BibitemOpen
  \bibfield  {author} {\bibinfo {author} {\bibfnamefont {B.}~\bibnamefont
  {Shan}}, \bibinfo {author} {\bibfnamefont {L.}~\bibnamefont {Huang}},
  \bibinfo {author} {\bibfnamefont {Y.}~\bibnamefont {Yang}}, \bibinfo {author}
  {\bibfnamefont {Y.}~\bibnamefont {Zhao}}, \bibinfo {author} {\bibfnamefont
  {J.}~\bibnamefont {Shen}}, \bibinfo {author} {\bibfnamefont {Z.}~\bibnamefont
  {Ye}}, \bibinfo {author} {\bibfnamefont {L.}~\bibnamefont {Chen}}, \bibinfo
  {author} {\bibfnamefont {Z.}~\bibnamefont {Meng}}, \bibinfo {author}
  {\bibfnamefont {P.}~\bibnamefont {Wang}}, \bibinfo {author} {\bibfnamefont
  {W.}~\bibnamefont {Han}},\ and\ \bibinfo {author} {\bibfnamefont
  {J.}~\bibnamefont {Zhang}},\ }\bibfield  {title} {\bibinfo {title}
  {Experimental study of magnetically insensitive transitions in an ultracold
  fermi gas of $^{40}\mathrm{K}$},\ }\href {https://doi.org/10.1103/jk8m-snjw}
  {\bibfield  {journal} {\bibinfo  {journal} {Phys. Rev. A}\ }\textbf {\bibinfo
  {volume} {113}},\ \bibinfo {pages} {023306} (\bibinfo {year}
  {2026})}\BibitemShut {NoStop}%
\bibitem [{\citenamefont {Breit}\ and\ \citenamefont
  {Rabi}(1931)}]{Rabi1931PR}%
  \BibitemOpen
  \bibfield  {author} {\bibinfo {author} {\bibfnamefont {G.}~\bibnamefont
  {Breit}}\ and\ \bibinfo {author} {\bibfnamefont {I.~I.}\ \bibnamefont
  {Rabi}},\ }\bibfield  {title} {\bibinfo {title} {Measurement of nuclear
  spin},\ }\href {https://doi.org/10.1103/PhysRev.38.2082.2} {\bibfield
  {journal} {\bibinfo  {journal} {Phys. Rev.}\ }\textbf {\bibinfo {volume}
  {38}},\ \bibinfo {pages} {2082} (\bibinfo {year} {1931})}\BibitemShut
  {NoStop}%
\bibitem [{\citenamefont {Steck}(2019)}]{Daniel2019K40}%
  \BibitemOpen
  \bibfield  {author} {\bibinfo {author} {\bibfnamefont {D.}~\bibnamefont
  {Steck}},\ }\href
  {https://www.tobiastiecke.nl/archive/PotassiumProperties.pdf} {\bibinfo
  {title} {{Alkali D Line Data for Potassium}}} (\bibinfo {year}
  {2019})\BibitemShut {NoStop}%
\bibitem [{SM2()}]{SM2026}%
  \BibitemOpen
  \href@noop {} {\bibinfo {title} {See {Supplemental Material} at \href{[URL
  will be inserted by publisher]}{[URL will be inserted by publisher]} for a
  detailed description of the magnetic-field-calibration-free theoretical
  model, experimental details and microwave pulse generation, {R}amsey
  measurement and fitting procedure, as well as the precise determination and
  uncertainty budget of the hyperfine constant {$A$}.}}\BibitemShut {Stop}%
\bibitem [{\citenamefont {Wei}\ \emph {et~al.}(2007{\natexlab{b}})\citenamefont
  {Wei}, \citenamefont {Xiong}, \citenamefont {Chen}, \citenamefont {Wang},
  \citenamefont {Guo},\ and\ \citenamefont {Zhang}}]{WeiDong2007CPL}%
  \BibitemOpen
  \bibfield  {author} {\bibinfo {author} {\bibfnamefont {D.}~\bibnamefont
  {Wei}}, \bibinfo {author} {\bibfnamefont {D.}~\bibnamefont {Xiong}}, \bibinfo
  {author} {\bibfnamefont {H.}~\bibnamefont {Chen}}, \bibinfo {author}
  {\bibfnamefont {P.}~\bibnamefont {Wang}}, \bibinfo {author} {\bibfnamefont
  {L.}~\bibnamefont {Guo}},\ and\ \bibinfo {author} {\bibfnamefont
  {J.}~\bibnamefont {Zhang}},\ }\bibfield  {title} {\bibinfo {title}
  {{Simultaneous Magneto-Optical Trapping of Fermionic $^{40}$K and Bosonic
  $^{87}$Rb Atoms}},\ }\href {https://doi.org/10.1088/0256-307X/24/6/030}
  {\bibfield  {journal} {\bibinfo  {journal} {Chin. Phys. Lett.}\ }\textbf
  {\bibinfo {volume} {24}},\ \bibinfo {pages} {1541} (\bibinfo {year}
  {2007}{\natexlab{b}})}\BibitemShut {NoStop}%
\bibitem [{\citenamefont {Chai}\ \emph {et~al.}(2012)\citenamefont {Chai},
  \citenamefont {Wei}, \citenamefont {Fu}, \citenamefont {Huang},\ and\
  \citenamefont {Zhang}}]{chaiShijie2012ActaSinQuantumOpt}%
  \BibitemOpen
  \bibfield  {author} {\bibinfo {author} {\bibfnamefont {S.}~\bibnamefont
  {Chai}}, \bibinfo {author} {\bibfnamefont {D.}~\bibnamefont {Wei}}, \bibinfo
  {author} {\bibfnamefont {Z.}~\bibnamefont {Fu}}, \bibinfo {author}
  {\bibfnamefont {L.}~\bibnamefont {Huang}},\ and\ \bibinfo {author}
  {\bibfnamefont {J.}~\bibnamefont {Zhang}},\ }\bibfield  {title} {\bibinfo
  {title} {{The design of a dipole traps for Bose-Einstein condensate and
  degenerate Fermi gas}},\ }\href {https://doi.org/10.3788/jqo20121802.0171}
  {\bibfield  {journal} {\bibinfo  {journal} {Acta Sin. Quantum Opt.}\ }\textbf
  {\bibinfo {volume} {18}},\ \bibinfo {pages} {171} (\bibinfo {year}
  {2012})}\BibitemShut {NoStop}%
\bibitem [{\citenamefont {Wei}\ \emph {et~al.}(2006)\citenamefont {Wei},
  \citenamefont {Xiong}, \citenamefont {Chen},\ and\ \citenamefont
  {Zhang}}]{Weidong2006ActaPhysSin}%
  \BibitemOpen
  \bibfield  {author} {\bibinfo {author} {\bibfnamefont {D.}~\bibnamefont
  {Wei}}, \bibinfo {author} {\bibfnamefont {D.}~\bibnamefont {Xiong}}, \bibinfo
  {author} {\bibfnamefont {H.}~\bibnamefont {Chen}},\ and\ \bibinfo {author}
  {\bibfnamefont {J.}~\bibnamefont {Zhang}},\ }\bibfield  {title} {\bibinfo
  {title} {{A laser diode system for $^{40}$K-$^{87}$Rb atomic cooling}},\
  }\href {https://doi.org/10.7498/aps.55.6342} {\bibfield  {journal} {\bibinfo
  {journal} {Acta Phys. Sin}\ }\textbf {\bibinfo {volume} {55}},\ \bibinfo
  {pages} {6342} (\bibinfo {year} {2006})}\BibitemShut {NoStop}%
\bibitem [{\citenamefont {Wang}\ \emph {et~al.}(2016)\citenamefont {Wang},
  \citenamefont {Chen}, \citenamefont {Wang},\ and\ \citenamefont
  {Zhang}}]{Wangaqiong2016ASQO}%
  \BibitemOpen
  \bibfield  {author} {\bibinfo {author} {\bibfnamefont {Y.}~\bibnamefont
  {Wang}}, \bibinfo {author} {\bibfnamefont {L.}~\bibnamefont {Chen}}, \bibinfo
  {author} {\bibfnamefont {P.}~\bibnamefont {Wang}},\ and\ \bibinfo {author}
  {\bibfnamefont {J.}~\bibnamefont {Zhang}},\ }\bibfield  {title} {\bibinfo
  {title} {{The Influence of Gravitys Effect on Loading Cold $^{87}$Rb Atoms
  into Quadrupole Magnetic Trap}},\ }\href
  {https://doi.org/10.3788/JQO20162201.0008} {\bibfield  {journal} {\bibinfo
  {journal} {{J. Quant. Opt.}}\ }\textbf {\bibinfo {volume} {22}},\ \bibinfo
  {pages} {50} (\bibinfo {year} {2016})}\BibitemShut {NoStop}%
\bibitem [{\citenamefont {Chen}\ \emph {et~al.}(2017)\citenamefont {Chen},
  \citenamefont {Yang}, \citenamefont {Meng}, \citenamefont {Huang},
  \citenamefont {Wang},\ and\ \citenamefont {Zhang}}]{liangchao2017ASQO}%
  \BibitemOpen
  \bibfield  {author} {\bibinfo {author} {\bibfnamefont {L.}~\bibnamefont
  {Chen}}, \bibinfo {author} {\bibfnamefont {G.}~\bibnamefont {Yang}}, \bibinfo
  {author} {\bibfnamefont {Z.}~\bibnamefont {Meng}}, \bibinfo {author}
  {\bibfnamefont {L.}~\bibnamefont {Huang}}, \bibinfo {author} {\bibfnamefont
  {P.}~\bibnamefont {Wang}},\ and\ \bibinfo {author} {\bibfnamefont
  {J.}~\bibnamefont {Zhang}},\ }\bibfield  {title} {\bibinfo {title}
  {{Electromagnetically Induced Transparency in $^{87}$Rb Bose-Einstein
  Condensate}},\ }\href {https://doi.org/10.3788/JQO20172303.0006} {\bibfield
  {journal} {\bibinfo  {journal} {{J. Quant. Opt.}}\ }\textbf {\bibinfo
  {volume} {23}},\ \bibinfo {pages} {246} (\bibinfo {year} {2017})}\BibitemShut
  {NoStop}%
\bibitem [{\citenamefont {Miao}\ \emph {et~al.}(2022)\citenamefont {Miao},
  \citenamefont {Bian}, \citenamefont {Shan}, \citenamefont {Chen},
  \citenamefont {Meng}, \citenamefont {Wang}, \citenamefont {Huang},\ and\
  \citenamefont {Zhang}}]{MiaoJie2022CPB}%
  \BibitemOpen
  \bibfield  {author} {\bibinfo {author} {\bibfnamefont {J.}~\bibnamefont
  {Miao}}, \bibinfo {author} {\bibfnamefont {G.}~\bibnamefont {Bian}}, \bibinfo
  {author} {\bibfnamefont {B.}~\bibnamefont {Shan}}, \bibinfo {author}
  {\bibfnamefont {L.}~\bibnamefont {Chen}}, \bibinfo {author} {\bibfnamefont
  {Z.}~\bibnamefont {Meng}}, \bibinfo {author} {\bibfnamefont {P.}~\bibnamefont
  {Wang}}, \bibinfo {author} {\bibfnamefont {L.}~\bibnamefont {Huang}},\ and\
  \bibinfo {author} {\bibfnamefont {J.}~\bibnamefont {Zhang}},\ }\bibfield
  {title} {\bibinfo {title} {{Achieving ultracold Bose--Fermi mixture of
  $^{87}$Rb and $^{40}$K with dual dark magnetic-optical-trap}},\ }\href
  {https://doi.org/10.1088/1674-1056/ac5882} {\bibfield  {journal} {\bibinfo
  {journal} {Chin. Phys. B}\ }\textbf {\bibinfo {volume} {31}},\ \bibinfo
  {pages} {080306} (\bibinfo {year} {2022})}\BibitemShut {NoStop}%
\bibitem [{\citenamefont {Ding}\ \emph {et~al.}(2024)\citenamefont {Ding},
  \citenamefont {Shan}, \citenamefont {Zhao}, \citenamefont {Yang},
  \citenamefont {Chen}, \citenamefont {Meng}, \citenamefont {Wang},
  \citenamefont {Huang},\ and\ \citenamefont {Zhang}}]{Ding2024CPB}%
  \BibitemOpen
  \bibfield  {author} {\bibinfo {author} {\bibfnamefont {P.}~\bibnamefont
  {Ding}}, \bibinfo {author} {\bibfnamefont {B.}~\bibnamefont {Shan}}, \bibinfo
  {author} {\bibfnamefont {Y.}~\bibnamefont {Zhao}}, \bibinfo {author}
  {\bibfnamefont {Y.}~\bibnamefont {Yang}}, \bibinfo {author} {\bibfnamefont
  {L.}~\bibnamefont {Chen}}, \bibinfo {author} {\bibfnamefont {Z.}~\bibnamefont
  {Meng}}, \bibinfo {author} {\bibfnamefont {P.}~\bibnamefont {Wang}}, \bibinfo
  {author} {\bibfnamefont {L.}~\bibnamefont {Huang}},\ and\ \bibinfo {author}
  {\bibfnamefont {J.}~\bibnamefont {Zhang}},\ }\bibfield  {title} {\bibinfo
  {title} {{Optimal preparation of Bose and Fermi atomic gas mixtures for
  $^{87}$Rb and $^{40}$K in a crossed optical dipole trap}},\ }\href
  {https://doi.org/10.1088/1674-1056/ad334d} {\bibfield  {journal} {\bibinfo
  {journal} {Chin. Phys. B}\ }\textbf {\bibinfo {volume} {33}},\ \bibinfo
  {pages} {063402} (\bibinfo {year} {2024})}\BibitemShut {NoStop}%
\bibitem [{\citenamefont {Peng}\ \emph {et~al.}(2018)\citenamefont {Peng},
  \citenamefont {Huang}, \citenamefont {Li}, \citenamefont {Meng},
  \citenamefont {Wang},\ and\ \citenamefont {Zhang}}]{Peng2018CPL}%
  \BibitemOpen
  \bibfield  {author} {\bibinfo {author} {\bibfnamefont {P.}~\bibnamefont
  {Peng}}, \bibinfo {author} {\bibfnamefont {L.-H.}\ \bibnamefont {Huang}},
  \bibinfo {author} {\bibfnamefont {D.-H.}\ \bibnamefont {Li}}, \bibinfo
  {author} {\bibfnamefont {Z.-M.}\ \bibnamefont {Meng}}, \bibinfo {author}
  {\bibfnamefont {P.-J.}\ \bibnamefont {Wang}},\ and\ \bibinfo {author}
  {\bibfnamefont {J.}~\bibnamefont {Zhang}},\ }\bibfield  {title} {\bibinfo
  {title} {Experimental {{Observation}} of {{Spin-Exchange}} in {{Ultracold
  Fermi Gases}}},\ }\href {https://doi.org/10.1088/0256-307X/35/3/033401}
  {\bibfield  {journal} {\bibinfo  {journal} {Chin. Phys. Lett.}\ }\textbf
  {\bibinfo {volume} {35}},\ \bibinfo {pages} {033401} (\bibinfo {year}
  {2018})}\BibitemShut {NoStop}%
\bibitem [{\citenamefont {Lundblad}\ \emph {et~al.}(2010)\citenamefont
  {Lundblad}, \citenamefont {Schlosser},\ and\ \citenamefont
  {Porto}}]{Porto2010PRA}%
  \BibitemOpen
  \bibfield  {author} {\bibinfo {author} {\bibfnamefont {N.}~\bibnamefont
  {Lundblad}}, \bibinfo {author} {\bibfnamefont {M.}~\bibnamefont
  {Schlosser}},\ and\ \bibinfo {author} {\bibfnamefont {J.~V.}\ \bibnamefont
  {Porto}},\ }\bibfield  {title} {\bibinfo {title} {Experimental observation of
  magic-wavelength behavior of $^{87}\mathrm{Rb}$ atoms in an optical
  lattice},\ }\href {https://doi.org/10.1103/PhysRevA.81.031611} {\bibfield
  {journal} {\bibinfo  {journal} {Phys. Rev. A}\ }\textbf {\bibinfo {volume}
  {81}},\ \bibinfo {pages} {031611} (\bibinfo {year} {2010})}\BibitemShut
  {NoStop}%
\bibitem [{\citenamefont {Yang}\ \emph {et~al.}(2016)\citenamefont {Yang},
  \citenamefont {He}, \citenamefont {Guo}, \citenamefont {Xu}, \citenamefont
  {Wang}, \citenamefont {Sheng}, \citenamefont {Liu}, \citenamefont {Wang},
  \citenamefont {Derevianko},\ and\ \citenamefont {Zhan}}]{Zhan2016PRL}%
  \BibitemOpen
  \bibfield  {author} {\bibinfo {author} {\bibfnamefont {J.}~\bibnamefont
  {Yang}}, \bibinfo {author} {\bibfnamefont {X.}~\bibnamefont {He}}, \bibinfo
  {author} {\bibfnamefont {R.}~\bibnamefont {Guo}}, \bibinfo {author}
  {\bibfnamefont {P.}~\bibnamefont {Xu}}, \bibinfo {author} {\bibfnamefont
  {K.}~\bibnamefont {Wang}}, \bibinfo {author} {\bibfnamefont {C.}~\bibnamefont
  {Sheng}}, \bibinfo {author} {\bibfnamefont {M.}~\bibnamefont {Liu}}, \bibinfo
  {author} {\bibfnamefont {J.}~\bibnamefont {Wang}}, \bibinfo {author}
  {\bibfnamefont {A.}~\bibnamefont {Derevianko}},\ and\ \bibinfo {author}
  {\bibfnamefont {M.}~\bibnamefont {Zhan}},\ }\bibfield  {title} {\bibinfo
  {title} {Coherence preservation of a single neutral atom qubit transferred
  between magic-intensity optical traps},\ }\href
  {https://doi.org/10.1103/PhysRevLett.117.123201} {\bibfield  {journal}
  {\bibinfo  {journal} {Phys. Rev. Lett.}\ }\textbf {\bibinfo {volume} {117}},\
  \bibinfo {pages} {123201} (\bibinfo {year} {2016})}\BibitemShut {NoStop}%
\bibitem [{\citenamefont {Steck}(2022)}]{DSteck}%
  \BibitemOpen
  \bibfield  {author} {\bibinfo {author} {\bibfnamefont {D.}~\bibnamefont
  {Steck}},\ }\href
  {https://atomoptics.uoregon.edu/~dsteck/teaching/quantum-optics/} {\bibinfo
  {title} {{Quantum and Atom Optics}}} (\bibinfo {year} {2022})\BibitemShut
  {NoStop}%
\bibitem [{\citenamefont {Gupta}\ \emph {et~al.}(2003)\citenamefont {Gupta},
  \citenamefont {Hadzibabic}, \citenamefont {Zwierlein}, \citenamefont {Stan},
  \citenamefont {Dieckmann}, \citenamefont {Schunck}, \citenamefont {van
  Kempen}, \citenamefont {Verhaar},\ and\ \citenamefont
  {Ketterle}}]{Gupta2003Science}%
  \BibitemOpen
  \bibfield  {author} {\bibinfo {author} {\bibfnamefont {S.}~\bibnamefont
  {Gupta}}, \bibinfo {author} {\bibfnamefont {Z.}~\bibnamefont {Hadzibabic}},
  \bibinfo {author} {\bibfnamefont {M.~W.}\ \bibnamefont {Zwierlein}}, \bibinfo
  {author} {\bibfnamefont {C.~A.}\ \bibnamefont {Stan}}, \bibinfo {author}
  {\bibfnamefont {K.}~\bibnamefont {Dieckmann}}, \bibinfo {author}
  {\bibfnamefont {C.~H.}\ \bibnamefont {Schunck}}, \bibinfo {author}
  {\bibfnamefont {E.~G.~M.}\ \bibnamefont {van Kempen}}, \bibinfo {author}
  {\bibfnamefont {B.~J.}\ \bibnamefont {Verhaar}},\ and\ \bibinfo {author}
  {\bibfnamefont {W.}~\bibnamefont {Ketterle}},\ }\bibfield  {title} {\bibinfo
  {title} {Radio-frequency spectroscopy of ultracold fermions},\ }\href
  {https://doi.org/10.1126/science.1085335} {\bibfield  {journal} {\bibinfo
  {journal} {Science}\ }\textbf {\bibinfo {volume} {300}},\ \bibinfo {pages}
  {1723} (\bibinfo {year} {2003})}\BibitemShut {NoStop}%
\bibitem [{\citenamefont {Zwierlein}\ \emph {et~al.}(2003)\citenamefont
  {Zwierlein}, \citenamefont {Hadzibabic}, \citenamefont {Gupta},\ and\
  \citenamefont {Ketterle}}]{Zwierlein2003PRL}%
  \BibitemOpen
  \bibfield  {author} {\bibinfo {author} {\bibfnamefont {M.~W.}\ \bibnamefont
  {Zwierlein}}, \bibinfo {author} {\bibfnamefont {Z.}~\bibnamefont
  {Hadzibabic}}, \bibinfo {author} {\bibfnamefont {S.}~\bibnamefont {Gupta}},\
  and\ \bibinfo {author} {\bibfnamefont {W.}~\bibnamefont {Ketterle}},\
  }\bibfield  {title} {\bibinfo {title} {Spectroscopic insensitivity to cold
  collisions in a two-state mixture of fermions},\ }\href
  {https://doi.org/10.1103/PhysRevLett.91.250404} {\bibfield  {journal}
  {\bibinfo  {journal} {Phys. Rev. Lett.}\ }\textbf {\bibinfo {volume} {91}},\
  \bibinfo {pages} {250404} (\bibinfo {year} {2003})}\BibitemShut {NoStop}%
\bibitem [{\citenamefont {Regal}\ \emph {et~al.}(2004)\citenamefont {Regal},
  \citenamefont {Greiner},\ and\ \citenamefont {Jin}}]{Regal2004PRL}%
  \BibitemOpen
  \bibfield  {author} {\bibinfo {author} {\bibfnamefont {C.~A.}\ \bibnamefont
  {Regal}}, \bibinfo {author} {\bibfnamefont {M.}~\bibnamefont {Greiner}},\
  and\ \bibinfo {author} {\bibfnamefont {D.~S.}\ \bibnamefont {Jin}},\
  }\bibfield  {title} {\bibinfo {title} {{Observation of Resonance Condensation
  of Fermionic Atom Pairs}},\ }\href
  {https://doi.org/10.1103/PhysRevLett.92.040403} {\bibfield  {journal}
  {\bibinfo  {journal} {Phys. Rev. Lett.}\ }\textbf {\bibinfo {volume} {92}},\
  \bibinfo {pages} {040403} (\bibinfo {year} {2004})}\BibitemShut {NoStop}%
\bibitem [{\citenamefont {Chin}\ \emph {et~al.}(2010)\citenamefont {Chin},
  \citenamefont {Grimm}, \citenamefont {Julienne},\ and\ \citenamefont
  {Tiesinga}}]{Tiesinga2010RMP}%
  \BibitemOpen
  \bibfield  {author} {\bibinfo {author} {\bibfnamefont {C.}~\bibnamefont
  {Chin}}, \bibinfo {author} {\bibfnamefont {R.}~\bibnamefont {Grimm}},
  \bibinfo {author} {\bibfnamefont {P.~S.}\ \bibnamefont {Julienne}},\ and\
  \bibinfo {author} {\bibfnamefont {E.}~\bibnamefont {Tiesinga}},\ }\bibfield
  {title} {\bibinfo {title} {Feshbach resonances in ultracold gases},\ }\href
  {https://doi.org/10.1103/RevModPhys.82.1225} {\bibfield  {journal} {\bibinfo
  {journal} {Rev. Mod. Phys.}\ }\textbf {\bibinfo {volume} {82}},\ \bibinfo
  {pages} {1225} (\bibinfo {year} {2010})}\BibitemShut {NoStop}%
\bibitem [{\citenamefont {Wille}\ \emph
  {et~al.}(2008{\natexlab{b}})\citenamefont {Wille}, \citenamefont
  {Spiegelhalder}, \citenamefont {Kerner}, \citenamefont {Naik}, \citenamefont
  {Trenkwalder}, \citenamefont {Hendl}, \citenamefont {Schreck}, \citenamefont
  {Grimm}, \citenamefont {Tiecke}, \citenamefont {Walraven}, \citenamefont
  {Kokkelmans}, \citenamefont {Tiesinga},\ and\ \citenamefont
  {Julienne}}]{Wille2008PRL}%
  \BibitemOpen
  \bibfield  {author} {\bibinfo {author} {\bibfnamefont {E.}~\bibnamefont
  {Wille}}, \bibinfo {author} {\bibfnamefont {F.~M.}\ \bibnamefont
  {Spiegelhalder}}, \bibinfo {author} {\bibfnamefont {G.}~\bibnamefont
  {Kerner}}, \bibinfo {author} {\bibfnamefont {D.}~\bibnamefont {Naik}},
  \bibinfo {author} {\bibfnamefont {A.}~\bibnamefont {Trenkwalder}}, \bibinfo
  {author} {\bibfnamefont {G.}~\bibnamefont {Hendl}}, \bibinfo {author}
  {\bibfnamefont {F.}~\bibnamefont {Schreck}}, \bibinfo {author} {\bibfnamefont
  {R.}~\bibnamefont {Grimm}}, \bibinfo {author} {\bibfnamefont {T.~G.}\
  \bibnamefont {Tiecke}}, \bibinfo {author} {\bibfnamefont {J.~T.~M.}\
  \bibnamefont {Walraven}}, \bibinfo {author} {\bibfnamefont {S.~J. J. M.~F.}\
  \bibnamefont {Kokkelmans}}, \bibinfo {author} {\bibfnamefont
  {E.}~\bibnamefont {Tiesinga}},\ and\ \bibinfo {author} {\bibfnamefont
  {P.~S.}\ \bibnamefont {Julienne}},\ }\bibfield  {title} {\bibinfo {title}
  {{Exploring an ultracold Fermi-Fermi mixture: Interspecies Feshbach
  resonances and scattering properties of $^{6}$Li and $^{40}$K}},\ }\href
  {https://doi.org/10.1103/PhysRevLett.100.053201} {\bibfield  {journal}
  {\bibinfo  {journal} {Phys. Rev. Lett.}\ }\textbf {\bibinfo {volume} {100}},\
  \bibinfo {pages} {053201} (\bibinfo {year} {2008}{\natexlab{b}})}\BibitemShut
  {NoStop}%
\bibitem [{\citenamefont {Regal}\ and\ \citenamefont {Jin}(2003)}]{Jin2003PRL}%
  \BibitemOpen
  \bibfield  {author} {\bibinfo {author} {\bibfnamefont {C.~A.}\ \bibnamefont
  {Regal}}\ and\ \bibinfo {author} {\bibfnamefont {D.~S.}\ \bibnamefont
  {Jin}},\ }\bibfield  {title} {\bibinfo {title} {{Measurement of Positive and
  Negative Scattering Lengths in a Fermi Gas of Atoms}},\ }\href
  {https://doi.org/10.1103/PhysRevLett.90.230404} {\bibfield  {journal}
  {\bibinfo  {journal} {Phys. Rev. Lett.}\ }\textbf {\bibinfo {volume} {90}},\
  \bibinfo {pages} {230404} (\bibinfo {year} {2003})}\BibitemShut {NoStop}%
\bibitem [{\citenamefont {Jiang}\ and\ \citenamefont
  {Mitroy}(2013)}]{Mitroy2013PRA}%
  \BibitemOpen
  \bibfield  {author} {\bibinfo {author} {\bibfnamefont {J.}~\bibnamefont
  {Jiang}}\ and\ \bibinfo {author} {\bibfnamefont {J.}~\bibnamefont {Mitroy}},\
  }\bibfield  {title} {\bibinfo {title} {Hyperfine effects on potassium
  tune-out wavelengths and polarizabilities},\ }\href
  {https://doi.org/10.1103/PhysRevA.88.032505} {\bibfield  {journal} {\bibinfo
  {journal} {Phys. Rev. A}\ }\textbf {\bibinfo {volume} {88}},\ \bibinfo
  {pages} {032505} (\bibinfo {year} {2013})}\BibitemShut {NoStop}%
\bibitem [{\citenamefont {Schmidt}\ \emph {et~al.}(2016)\citenamefont
  {Schmidt}, \citenamefont {Mayer}, \citenamefont {Hohmann}, \citenamefont
  {Lausch}, \citenamefont {Kindermann},\ and\ \citenamefont
  {Widera}}]{Widera2016PRA}%
  \BibitemOpen
  \bibfield  {author} {\bibinfo {author} {\bibfnamefont {F.}~\bibnamefont
  {Schmidt}}, \bibinfo {author} {\bibfnamefont {D.}~\bibnamefont {Mayer}},
  \bibinfo {author} {\bibfnamefont {M.}~\bibnamefont {Hohmann}}, \bibinfo
  {author} {\bibfnamefont {T.}~\bibnamefont {Lausch}}, \bibinfo {author}
  {\bibfnamefont {F.}~\bibnamefont {Kindermann}},\ and\ \bibinfo {author}
  {\bibfnamefont {A.}~\bibnamefont {Widera}},\ }\bibfield  {title} {\bibinfo
  {title} {{Precision measurement of the $^{87}\text{Rb}$ tune-out wavelength
  in the hyperfine ground state $F=1$ at 790 nm}},\ }\href
  {https://doi.org/10.1103/PhysRevA.93.022507} {\bibfield  {journal} {\bibinfo
  {journal} {Phys. Rev. A}\ }\textbf {\bibinfo {volume} {93}},\ \bibinfo
  {pages} {022507} (\bibinfo {year} {2016})}\BibitemShut {NoStop}%
\bibitem [{\citenamefont {Li}\ \emph {et~al.}(2019)\citenamefont {Li},
  \citenamefont {Tian}, \citenamefont {Wu}, \citenamefont {Li}, \citenamefont
  {Li}, \citenamefont {Liu}, \citenamefont {Zhang},\ and\ \citenamefont
  {Zhang}}]{Tiancai2019PRL}%
  \BibitemOpen
  \bibfield  {author} {\bibinfo {author} {\bibfnamefont {G.}~\bibnamefont
  {Li}}, \bibinfo {author} {\bibfnamefont {Y.}~\bibnamefont {Tian}}, \bibinfo
  {author} {\bibfnamefont {W.}~\bibnamefont {Wu}}, \bibinfo {author}
  {\bibfnamefont {S.}~\bibnamefont {Li}}, \bibinfo {author} {\bibfnamefont
  {X.}~\bibnamefont {Li}}, \bibinfo {author} {\bibfnamefont {Y.}~\bibnamefont
  {Liu}}, \bibinfo {author} {\bibfnamefont {P.}~\bibnamefont {Zhang}},\ and\
  \bibinfo {author} {\bibfnamefont {T.}~\bibnamefont {Zhang}},\ }\bibfield
  {title} {\bibinfo {title} {Triply magic conditions for microwave transition
  of optically trapped alkali-metal atoms},\ }\href
  {https://doi.org/10.1103/PhysRevLett.123.253602} {\bibfield  {journal}
  {\bibinfo  {journal} {Phys. Rev. Lett.}\ }\textbf {\bibinfo {volume} {123}},\
  \bibinfo {pages} {253602} (\bibinfo {year} {2019})}\BibitemShut {NoStop}%
\bibitem [{\citenamefont {Wen}\ \emph {et~al.}(2021)\citenamefont {Wen},
  \citenamefont {Meng}, \citenamefont {Wang}, \citenamefont {Chen},
  \citenamefont {Huang}, \citenamefont {Wang},\ and\ \citenamefont
  {Zhang}}]{Wen2021JOSAB}%
  \BibitemOpen
  \bibfield  {author} {\bibinfo {author} {\bibfnamefont {K.}~\bibnamefont
  {Wen}}, \bibinfo {author} {\bibfnamefont {Z.}~\bibnamefont {Meng}}, \bibinfo
  {author} {\bibfnamefont {L.}~\bibnamefont {Wang}}, \bibinfo {author}
  {\bibfnamefont {L.}~\bibnamefont {Chen}}, \bibinfo {author} {\bibfnamefont
  {L.}~\bibnamefont {Huang}}, \bibinfo {author} {\bibfnamefont
  {P.}~\bibnamefont {Wang}},\ and\ \bibinfo {author} {\bibfnamefont
  {J.}~\bibnamefont {Zhang}},\ }\bibfield  {title} {\bibinfo {title}
  {{Experimental study of tune-out wavelengths for spin-dependent optical
  lattice in $^{87}$Rb Bose--Einstein condensation}},\ }\href
  {https://doi.org/10.1364/JOSAB.432448} {\bibfield  {journal} {\bibinfo
  {journal} {J. Opt. Soc. Am. B}\ }\textbf {\bibinfo {volume} {38}},\ \bibinfo
  {pages} {3269} (\bibinfo {year} {2021})}\BibitemShut {NoStop}%
\end{thebibliography}%


\begin{thebibliography}{15}%
\makeatletter
\providecommand \@ifxundefined [1]{%
 \@ifx{#1\undefined}
}%
\providecommand \@ifnum [1]{%
 \ifnum #1\expandafter \@firstoftwo
 \else \expandafter \@secondoftwo
 \fi
}%
\providecommand \@ifx [1]{%
 \ifx #1\expandafter \@firstoftwo
 \else \expandafter \@secondoftwo
 \fi
}%
\providecommand \natexlab [1]{#1}%
\providecommand \enquote  [1]{``#1''}%
\providecommand \bibnamefont  [1]{#1}%
\providecommand \bibfnamefont [1]{#1}%
\providecommand \citenamefont [1]{#1}%
\providecommand \href@noop [0]{\@secondoftwo}%
\providecommand \href [0]{\begingroup \@sanitize@url \@href}%
\providecommand \@href[1]{\@@startlink{#1}\@@href}%
\providecommand \@@href[1]{\endgroup#1\@@endlink}%
\providecommand \@sanitize@url [0]{\catcode `\\12\catcode `\$12\catcode
  `\&12\catcode `\#12\catcode `\^12\catcode `\_12\catcode `\%12\relax}%
\providecommand \@@startlink[1]{}%
\providecommand \@@endlink[0]{}%
\providecommand \url  [0]{\begingroup\@sanitize@url \@url }%
\providecommand \@url [1]{\endgroup\@href {#1}{\urlprefix }}%
\providecommand \urlprefix  [0]{URL }%
\providecommand \Eprint [0]{\href }%
\providecommand \doibase [0]{https://doi.org/}%
\providecommand \selectlanguage [0]{\@gobble}%
\providecommand \bibinfo  [0]{\@secondoftwo}%
\providecommand \bibfield  [0]{\@secondoftwo}%
\providecommand \translation [1]{[#1]}%
\providecommand \BibitemOpen [0]{}%
\providecommand \bibitemStop [0]{}%
\providecommand \bibitemNoStop [0]{.\EOS\space}%
\providecommand \EOS [0]{\spacefactor3000\relax}%
\providecommand \BibitemShut  [1]{\csname bibitem#1\endcsname}%
\let\auto@bib@innerbib\@empty
%</preamble>
\bibitem [{\citenamefont {Shan}\ \emph {et~al.}(2026)\citenamefont {Shan},
  \citenamefont {Huang}, \citenamefont {Yang}, \citenamefont {Zhao},
  \citenamefont {Shen}, \citenamefont {Ye}, \citenamefont {Chen}, \citenamefont
  {Meng}, \citenamefont {Wang}, \citenamefont {Han},\ and\ \citenamefont
  {Zhang}}]{Shan2026PRA}%
  \BibitemOpen
  \bibfield  {author} {\bibinfo {author} {\bibfnamefont {B.}~\bibnamefont
  {Shan}}, \bibinfo {author} {\bibfnamefont {L.}~\bibnamefont {Huang}},
  \bibinfo {author} {\bibfnamefont {Y.}~\bibnamefont {Yang}}, \bibinfo {author}
  {\bibfnamefont {Y.}~\bibnamefont {Zhao}}, \bibinfo {author} {\bibfnamefont
  {J.}~\bibnamefont {Shen}}, \bibinfo {author} {\bibfnamefont {Z.}~\bibnamefont
  {Ye}}, \bibinfo {author} {\bibfnamefont {L.}~\bibnamefont {Chen}}, \bibinfo
  {author} {\bibfnamefont {Z.}~\bibnamefont {Meng}}, \bibinfo {author}
  {\bibfnamefont {P.}~\bibnamefont {Wang}}, \bibinfo {author} {\bibfnamefont
  {W.}~\bibnamefont {Han}},\ and\ \bibinfo {author} {\bibfnamefont
  {J.}~\bibnamefont {Zhang}},\ }\href {https://doi.org/10.1103/jk8m-snjw}
  {\bibfield  {journal} {\bibinfo  {journal} {Phys. Rev. A}\ }\textbf {\bibinfo
  {volume} {113}},\ \bibinfo {pages} {023306} (\bibinfo {year}
  {2026})}\BibitemShut {NoStop}%
\bibitem [{\citenamefont {Breit}\ and\ \citenamefont
  {Rabi}(1931)}]{Rabi1931PR}%
  \BibitemOpen
  \bibfield  {author} {\bibinfo {author} {\bibfnamefont {G.}~\bibnamefont
  {Breit}}\ and\ \bibinfo {author} {\bibfnamefont {I.~I.}\ \bibnamefont
  {Rabi}},\ }\href {https://doi.org/10.1103/PhysRev.38.2082.2} {\bibfield
  {journal} {\bibinfo  {journal} {Phys. Rev.}\ }\textbf {\bibinfo {volume}
  {38}},\ \bibinfo {pages} {2082} (\bibinfo {year} {1931})}\BibitemShut
  {NoStop}%
\bibitem [{\citenamefont {Steck}(2019)}]{Daniel2019K40}%
  \BibitemOpen
  \bibfield  {author} {\bibinfo {author} {\bibfnamefont {D.}~\bibnamefont
  {Steck}},\ }\href
  {https://www.tobiastiecke.nl/archive/PotassiumProperties.pdf} {\bibinfo
  {title} {{Alkali D Line Data for Potassium}}} (\bibinfo {year}
  {2019})\BibitemShut {NoStop}%
\bibitem [{\citenamefont {Arimondo}\ \emph {et~al.}(1977)\citenamefont
  {Arimondo}, \citenamefont {Inguscio},\ and\ \citenamefont
  {Violino}}]{Arimondo1977RMP}%
  \BibitemOpen
  \bibfield  {author} {\bibinfo {author} {\bibfnamefont {E.}~\bibnamefont
  {Arimondo}}, \bibinfo {author} {\bibfnamefont {M.}~\bibnamefont {Inguscio}},\
  and\ \bibinfo {author} {\bibfnamefont {P.}~\bibnamefont {Violino}},\ }\href
  {https://doi.org/10.1103/RevModPhys.49.31} {\bibfield  {journal} {\bibinfo
  {journal} {Rev. Mod. Phys.}\ }\textbf {\bibinfo {volume} {49}},\ \bibinfo
  {pages} {31} (\bibinfo {year} {1977})}\BibitemShut {NoStop}%
\bibitem [{\citenamefont {Peng}\ \emph {et~al.}(2018)\citenamefont {Peng},
  \citenamefont {Huang}, \citenamefont {Li}, \citenamefont {Meng},
  \citenamefont {Wang},\ and\ \citenamefont {Zhang}}]{Peng2018CPL}%
  \BibitemOpen
  \bibfield  {author} {\bibinfo {author} {\bibfnamefont {P.}~\bibnamefont
  {Peng}}, \bibinfo {author} {\bibfnamefont {L.-H.}\ \bibnamefont {Huang}},
  \bibinfo {author} {\bibfnamefont {D.-H.}\ \bibnamefont {Li}}, \bibinfo
  {author} {\bibfnamefont {Z.-M.}\ \bibnamefont {Meng}}, \bibinfo {author}
  {\bibfnamefont {P.-J.}\ \bibnamefont {Wang}},\ and\ \bibinfo {author}
  {\bibfnamefont {J.}~\bibnamefont {Zhang}},\ }\href
  {https://doi.org/10.1088/0256-307X/35/3/033401} {\bibfield  {journal}
  {\bibinfo  {journal} {Chin. Phys. Lett.}\ }\textbf {\bibinfo {volume} {35}},\
  \bibinfo {pages} {033401} (\bibinfo {year} {2018})}\BibitemShut {NoStop}%
\bibitem [{\citenamefont {Li}\ \emph {et~al.}(2023)\citenamefont {Li},
  \citenamefont {Gu}, \citenamefont {Shi}, \citenamefont {Wang},\ and\
  \citenamefont {Zhang}}]{Ziliang2023CPB}%
  \BibitemOpen
  \bibfield  {author} {\bibinfo {author} {\bibfnamefont {Z.}~\bibnamefont
  {Li}}, \bibinfo {author} {\bibfnamefont {Z.}~\bibnamefont {Gu}}, \bibinfo
  {author} {\bibfnamefont {Z.}~\bibnamefont {Shi}}, \bibinfo {author}
  {\bibfnamefont {P.}~\bibnamefont {Wang}},\ and\ \bibinfo {author}
  {\bibfnamefont {J.}~\bibnamefont {Zhang}},\ }\href
  {https://doi.org/10.1088/1674-1056/aca14f} {\bibfield  {journal} {\bibinfo
  {journal} {Chin. Phys. B}\ }\textbf {\bibinfo {volume} {32}},\ \bibinfo
  {pages} {023701} (\bibinfo {year} {2023})}\BibitemShut {NoStop}%
\bibitem [{\citenamefont {Graham}\ \emph {et~al.}(2022)\citenamefont {Graham},
  \citenamefont {Song}, \citenamefont {Scott}, \citenamefont {Poole},
  \citenamefont {Phuttitarn}, \citenamefont {Jooya}, \citenamefont {Eichler},
  \citenamefont {Jiang}, \citenamefont {Marra}, \citenamefont {Grinkemeyer},
  \citenamefont {Kwon}, \citenamefont {Ebert}, \citenamefont {Cherek},
  \citenamefont {Lichtman}, \citenamefont {Gillette}, \citenamefont {Gilbert},
  \citenamefont {Bowman}, \citenamefont {Ballance}, \citenamefont {Campbell},
  \citenamefont {Dahl}, \citenamefont {Crawford}, \citenamefont {Blunt},
  \citenamefont {Rogers}, \citenamefont {Noel},\ and\ \citenamefont
  {Saffman}}]{Graham2022Nature}%
  \BibitemOpen
  \bibfield  {author} {\bibinfo {author} {\bibfnamefont {T.~M.}\ \bibnamefont
  {Graham}}, \bibinfo {author} {\bibfnamefont {Y.}~\bibnamefont {Song}},
  \bibinfo {author} {\bibfnamefont {J.}~\bibnamefont {Scott}}, \bibinfo
  {author} {\bibfnamefont {C.}~\bibnamefont {Poole}}, \bibinfo {author}
  {\bibfnamefont {L.}~\bibnamefont {Phuttitarn}}, \bibinfo {author}
  {\bibfnamefont {K.}~\bibnamefont {Jooya}}, \bibinfo {author} {\bibfnamefont
  {P.}~\bibnamefont {Eichler}}, \bibinfo {author} {\bibfnamefont
  {X.}~\bibnamefont {Jiang}}, \bibinfo {author} {\bibfnamefont
  {A.}~\bibnamefont {Marra}}, \bibinfo {author} {\bibfnamefont
  {B.}~\bibnamefont {Grinkemeyer}}, \bibinfo {author} {\bibfnamefont
  {M.}~\bibnamefont {Kwon}}, \bibinfo {author} {\bibfnamefont {M.}~\bibnamefont
  {Ebert}}, \bibinfo {author} {\bibfnamefont {J.}~\bibnamefont {Cherek}},
  \bibinfo {author} {\bibfnamefont {M.~T.}\ \bibnamefont {Lichtman}}, \bibinfo
  {author} {\bibfnamefont {M.}~\bibnamefont {Gillette}}, \bibinfo {author}
  {\bibfnamefont {J.}~\bibnamefont {Gilbert}}, \bibinfo {author} {\bibfnamefont
  {D.}~\bibnamefont {Bowman}}, \bibinfo {author} {\bibfnamefont
  {T.}~\bibnamefont {Ballance}}, \bibinfo {author} {\bibfnamefont
  {C.}~\bibnamefont {Campbell}}, \bibinfo {author} {\bibfnamefont {E.~D.}\
  \bibnamefont {Dahl}}, \bibinfo {author} {\bibfnamefont {O.}~\bibnamefont
  {Crawford}}, \bibinfo {author} {\bibfnamefont {N.~S.}\ \bibnamefont {Blunt}},
  \bibinfo {author} {\bibfnamefont {B.}~\bibnamefont {Rogers}}, \bibinfo
  {author} {\bibfnamefont {T.}~\bibnamefont {Noel}},\ and\ \bibinfo {author}
  {\bibfnamefont {M.}~\bibnamefont {Saffman}},\ }\href
  {https://doi.org/10.1038/s41586-022-04603-6} {\bibfield  {journal} {\bibinfo
  {journal} {Nature}\ }\textbf {\bibinfo {volume} {604}},\ \bibinfo {pages}
  {457} (\bibinfo {year} {2022})}\BibitemShut {NoStop}%
\bibitem [{\citenamefont {Ivannikov}\ and\ \citenamefont
  {Sidorov}(2018)}]{Ivannikov2018JPB}%
  \BibitemOpen
  \bibfield  {author} {\bibinfo {author} {\bibfnamefont {V.}~\bibnamefont
  {Ivannikov}}\ and\ \bibinfo {author} {\bibfnamefont {A.~I.}\ \bibnamefont
  {Sidorov}},\ }\href {https://doi.org/10.1088/1361-6455/aadfca} {\bibfield
  {journal} {\bibinfo  {journal} {J. Phys. B: At. Mol. Opt. Phys.}\ }\textbf
  {\bibinfo {volume} {51}},\ \bibinfo {pages} {205002} (\bibinfo {year}
  {2018})}\BibitemShut {NoStop}%
\bibitem [{\citenamefont {Xu}\ \emph {et~al.}(2019)\citenamefont {Xu},
  \citenamefont {Wang}, \citenamefont {Jiao}, \citenamefont {Yi}, \citenamefont
  {Sun},\ and\ \citenamefont {Chen}}]{Xu2019RSI}%
  \BibitemOpen
  \bibfield  {author} {\bibinfo {author} {\bibfnamefont {X.-T.}\ \bibnamefont
  {Xu}}, \bibinfo {author} {\bibfnamefont {Z.-Y.}\ \bibnamefont {Wang}},
  \bibinfo {author} {\bibfnamefont {R.-H.}\ \bibnamefont {Jiao}}, \bibinfo
  {author} {\bibfnamefont {C.-R.}\ \bibnamefont {Yi}}, \bibinfo {author}
  {\bibfnamefont {W.}~\bibnamefont {Sun}},\ and\ \bibinfo {author}
  {\bibfnamefont {S.}~\bibnamefont {Chen}},\ }\href
  {https://doi.org/10.1063/1.5087957} {\bibfield  {journal} {\bibinfo
  {journal} {Rev. Sci. Instrum.}\ }\textbf {\bibinfo {volume} {90}},\ \bibinfo
  {pages} {054708} (\bibinfo {year} {2019})}\BibitemShut {NoStop}%
\bibitem [{\citenamefont {H\"olzl}\ \emph {et~al.}(2024)\citenamefont
  {H\"olzl}, \citenamefont {G\"otzelmann}, \citenamefont {Pultinevicius},
  \citenamefont {Wirth},\ and\ \citenamefont {Meinert}}]{Meinert2024PRX}%
  \BibitemOpen
  \bibfield  {author} {\bibinfo {author} {\bibfnamefont {C.}~\bibnamefont
  {H\"olzl}}, \bibinfo {author} {\bibfnamefont {A.}~\bibnamefont
  {G\"otzelmann}}, \bibinfo {author} {\bibfnamefont {E.}~\bibnamefont
  {Pultinevicius}}, \bibinfo {author} {\bibfnamefont {M.}~\bibnamefont
  {Wirth}},\ and\ \bibinfo {author} {\bibfnamefont {F.}~\bibnamefont
  {Meinert}},\ }\href {https://doi.org/10.1103/PhysRevX.14.021024} {\bibfield
  {journal} {\bibinfo  {journal} {Phys. Rev. X}\ }\textbf {\bibinfo {volume}
  {14}},\ \bibinfo {pages} {021024} (\bibinfo {year} {2024})}\BibitemShut
  {NoStop}%
\bibitem [{\citenamefont {Tiengo}\ \emph {et~al.}(2025)\citenamefont {Tiengo},
  \citenamefont {Eid}, \citenamefont {Apfel}, \citenamefont {Brulin},\ and\
  \citenamefont {Bourdel}}]{Bourdel2025RSI}%
  \BibitemOpen
  \bibfield  {author} {\bibinfo {author} {\bibfnamefont {S.}~\bibnamefont
  {Tiengo}}, \bibinfo {author} {\bibfnamefont {R.}~\bibnamefont {Eid}},
  \bibinfo {author} {\bibfnamefont {M.}~\bibnamefont {Apfel}}, \bibinfo
  {author} {\bibfnamefont {G.}~\bibnamefont {Brulin}},\ and\ \bibinfo {author}
  {\bibfnamefont {T.}~\bibnamefont {Bourdel}},\ }\href
  {https://doi.org/10.1063/5.0258855} {\bibfield  {journal} {\bibinfo
  {journal} {Rev. Sci. Instrum.}\ }\textbf {\bibinfo {volume} {96}},\ \bibinfo
  {pages} {063201} (\bibinfo {year} {2025})}\BibitemShut {NoStop}%
\bibitem [{\citenamefont {Steck}(2022)}]{DSteck}%
  \BibitemOpen
  \bibfield  {author} {\bibinfo {author} {\bibfnamefont {D.}~\bibnamefont
  {Steck}},\ }\href
  {https://atomoptics.uoregon.edu/~dsteck/teaching/quantum-optics/} {\bibinfo
  {title} {{Quantum and Atom Optics}}} (\bibinfo {year} {2022})\BibitemShut
  {NoStop}%
\bibitem [{\citenamefont {Yang}\ \emph {et~al.}(2016)\citenamefont {Yang},
  \citenamefont {He}, \citenamefont {Guo}, \citenamefont {Xu}, \citenamefont
  {Wang}, \citenamefont {Sheng}, \citenamefont {Liu}, \citenamefont {Wang},
  \citenamefont {Derevianko},\ and\ \citenamefont {Zhan}}]{Zhan2016PRL}%
  \BibitemOpen
  \bibfield  {author} {\bibinfo {author} {\bibfnamefont {J.}~\bibnamefont
  {Yang}}, \bibinfo {author} {\bibfnamefont {X.}~\bibnamefont {He}}, \bibinfo
  {author} {\bibfnamefont {R.}~\bibnamefont {Guo}}, \bibinfo {author}
  {\bibfnamefont {P.}~\bibnamefont {Xu}}, \bibinfo {author} {\bibfnamefont
  {K.}~\bibnamefont {Wang}}, \bibinfo {author} {\bibfnamefont {C.}~\bibnamefont
  {Sheng}}, \bibinfo {author} {\bibfnamefont {M.}~\bibnamefont {Liu}}, \bibinfo
  {author} {\bibfnamefont {J.}~\bibnamefont {Wang}}, \bibinfo {author}
  {\bibfnamefont {A.}~\bibnamefont {Derevianko}},\ and\ \bibinfo {author}
  {\bibfnamefont {M.}~\bibnamefont {Zhan}},\ }\href
  {https://doi.org/10.1103/PhysRevLett.117.123201} {\bibfield  {journal}
  {\bibinfo  {journal} {Phys. Rev. Lett.}\ }\textbf {\bibinfo {volume} {117}},\
  \bibinfo {pages} {123201} (\bibinfo {year} {2016})}\BibitemShut {NoStop}%
\bibitem [{\citenamefont {Gupta}\ \emph {et~al.}(2003)\citenamefont {Gupta},
  \citenamefont {Hadzibabic}, \citenamefont {Zwierlein}, \citenamefont {Stan},
  \citenamefont {Dieckmann}, \citenamefont {Schunck}, \citenamefont {Verhaar},\
  and\ \citenamefont {Ketterle}}]{Ketterle2003Science}%
  \BibitemOpen
  \bibfield  {author} {\bibinfo {author} {\bibfnamefont {S.}~\bibnamefont
  {Gupta}}, \bibinfo {author} {\bibfnamefont {Z.}~\bibnamefont {Hadzibabic}},
  \bibinfo {author} {\bibfnamefont {M.~W.}\ \bibnamefont {Zwierlein}}, \bibinfo
  {author} {\bibfnamefont {C.~A.}\ \bibnamefont {Stan}}, \bibinfo {author}
  {\bibfnamefont {K.}~\bibnamefont {Dieckmann}}, \bibinfo {author}
  {\bibfnamefont {C.~H.}\ \bibnamefont {Schunck}}, \bibinfo {author}
  {\bibfnamefont {B.~J.}\ \bibnamefont {Verhaar}},\ and\ \bibinfo {author}
  {\bibfnamefont {W.}~\bibnamefont {Ketterle}},\ }\href
  {https://doi.org/10.1126/science.1085335} {\bibfield  {journal} {\bibinfo
  {journal} {Science}\ }\textbf {\bibinfo {volume} {300}},\ \bibinfo {pages}
  {1723} (\bibinfo {year} {2003})}\BibitemShut {NoStop}%
\bibitem [{\citenamefont {Zwierlein}\ \emph {et~al.}(2003)\citenamefont
  {Zwierlein}, \citenamefont {Hadzibabic}, \citenamefont {Gupta},\ and\
  \citenamefont {Ketterle}}]{Zwierlein2003PRL}%
  \BibitemOpen
  \bibfield  {author} {\bibinfo {author} {\bibfnamefont {M.~W.}\ \bibnamefont
  {Zwierlein}}, \bibinfo {author} {\bibfnamefont {Z.}~\bibnamefont
  {Hadzibabic}}, \bibinfo {author} {\bibfnamefont {S.}~\bibnamefont {Gupta}},\
  and\ \bibinfo {author} {\bibfnamefont {W.}~\bibnamefont {Ketterle}},\ }\href
  {https://doi.org/10.1103/PhysRevLett.91.250404} {\bibfield  {journal}
  {\bibinfo  {journal} {Phys. Rev. Lett.}\ }\textbf {\bibinfo {volume} {91}},\
  \bibinfo {pages} {250404} (\bibinfo {year} {2003})}\BibitemShut {NoStop}%
\end{thebibliography}%

\end{document}

\title{Supplemental Material for ``Magnetic-Field-Calibration-Free Determination of the Hyperfine Constant $A$ in Ultracold Fermi gases of $^{40}$K"}

\author{Yajing Yang}
\thanks{These authors contributed equally to this work.}
\affiliation{State Key Laboratory of Quantum Optics Technologies and Devices, \\  Institute of Opto-electronics, Collaborative Innovation Center of Extreme Optics, Shanxi University, Taiyuan, Shanxi 030006, China}
\author{Biao Shan}
\thanks{These authors contributed equally to this work.}
\affiliation{State Key Laboratory of Quantum Optics Technologies and Devices, \\  Institute of Opto-electronics, Collaborative Innovation Center of Extreme Optics, Shanxi University, Taiyuan, Shanxi 030006, China}
\author{Yuhang Zhao}
\affiliation{State Key Laboratory of Quantum Optics Technologies and Devices, \\  Institute of Opto-electronics, Collaborative Innovation Center of Extreme Optics, Shanxi University, Taiyuan, Shanxi 030006, China}
\author{Jiahui Shen}
\affiliation{State Key Laboratory of Quantum Optics Technologies and Devices, \\  Institute of Opto-electronics, Collaborative Innovation Center of Extreme Optics, Shanxi University, Taiyuan, Shanxi 030006, China}
\author{Zhuxiong Ye}
\affiliation{State Key Laboratory of Quantum Optics Technologies and Devices, \\  Institute of Opto-electronics, Collaborative Innovation Center of Extreme Optics, Shanxi University, Taiyuan, Shanxi 030006, China}
\author{Liangchao Chen}
\affiliation{State Key Laboratory of Quantum Optics Technologies and Devices, \\  Institute of Opto-electronics, Collaborative Innovation Center of Extreme Optics, Shanxi University, Taiyuan, Shanxi 030006, China}
\affiliation{Hefei National Laboratory, Hefei, Anhui 230088, China.}
\author{Zengming Meng}
\affiliation{State Key Laboratory of Quantum Optics Technologies and Devices, \\  Institute of Opto-electronics, Collaborative Innovation Center of Extreme Optics, Shanxi University, Taiyuan, Shanxi 030006, China}
\affiliation{Hefei National Laboratory, Hefei, Anhui 230088, China.}
\author{Pengjun Wang}
\affiliation{State Key Laboratory of Quantum Optics Technologies and Devices, \\  Institute of Opto-electronics, Collaborative Innovation Center of Extreme Optics, Shanxi University, Taiyuan, Shanxi 030006, China}
\affiliation{Hefei National Laboratory, Hefei, Anhui 230088, China.}
\author{Wei Han}
\email[Contact author:]{hanwei.irain@gmail.com;}
\affiliation{State Key Laboratory of Quantum Optics Technologies and Devices, \\  Institute of Opto-electronics, Collaborative Innovation Center of Extreme Optics, Shanxi University, Taiyuan, Shanxi 030006, China}
\affiliation{Hefei National Laboratory, Hefei, Anhui 230088, China.}
\author{Jing Zhang}
\email[Contact author:]{jzhang74@sxu.edu.cn;}
\affiliation{State Key Laboratory of Quantum Optics Technologies and Devices, \\  Institute of Opto-electronics, Collaborative Innovation Center of Extreme Optics, Shanxi University, Taiyuan, Shanxi 030006, China}
\affiliation{Hefei National Laboratory, Hefei, Anhui 230088, China.}
\author{Lianghui Huang}
\email[Contact author:]{huanglh06@sxu.edu.cn}
\affiliation{State Key Laboratory of Quantum Optics Technologies and Devices, \\  Institute of Opto-electronics, Collaborative Innovation Center of Extreme Optics, Shanxi University, Taiyuan, Shanxi 030006, China}
\affiliation{Hefei National Laboratory, Hefei, Anhui 230088, China.}

\date{\today}

\maketitle

\tableofcontents

\begin{widetext}
\setcounter{equation}{0}
\renewcommand{\theequation}{S\arabic{equation}}

\section*{I. Theoritical model for magnetic-field-calibration-free determination}

We consider two magnetically insensitive microwave single-photon transitions labeled $\alpha$ for $|F=9/2,m_F=-1/2\rangle$ $\Leftrightarrow$$|7/2, 1/2\rangle$ and $\beta$ for $|9/2, 1/2\rangle \Leftrightarrow |7/2, -1/2\rangle$ within the $^{40}\mathrm{K}$ ground-state hyperfine manifold~\cite{Shan2026PRA}. In the presence of an external magnetic field $B$, the Zeeman energies of the hyperfine states depend on the magnetic quantum number $m_{F}$ and are described by the Breit--Rabi formula~\cite{Rabi1931PR,Daniel2019K40}
\begin{equation}\label{Breit}
E_{F,m_F}(B)= -\frac{A}{4}+\mu_B g_I m_F B
\pm \frac{A(2I+1)}{4}\sqrt{1+\frac{4m_F}{2I+1}x+x^2},
\end{equation}
with
\begin{equation}
x=\frac{2(g_J-g_I)\mu_B}{A(2I+1)}B.
\end{equation}
Here $A$ is the hyperfine constant and $\mu_B$ is the Bohr magneton.
The parameters $g_J=2.00229421(24)$ and $g_I=0.000176490(34)$ are the electronic and nuclear Land\'{e} $g$ factors of the ground state, respectively~\cite{Arimondo1977RMP, Daniel2019K40}. For $^{40}$K, the nuclear spin is $I=4$, and the upper (lower) sign in ``$\pm$'' corresponds to $F=I+1/2$ ($F=I-1/2$).

The transition angular frequencies for the two magnetically insensitive microwave single-photon transitions are obtained from the relevant energy differences
\begin{align}
\omega_\alpha(B) &= \frac{E_{7/2,+1/2}(B)-E_{9/2,-1/2}(B)}{\hbar}, \label{alpha}\\
\omega_\beta(B)  &= \frac{E_{7/2,-1/2}(B)-E_{9/2,+1/2}(B)}{\hbar}. \label{beta}
\end{align}
By substituting Eq.~(\ref{Breit}) into Eq.~(\ref{alpha}) and Eq.~(\ref{beta}), we obtain a set of coupled equations
\begin{align}
\omega_{\alpha}(B)-\omega_{\beta}(B) &= 2\mu_{B}g_{I}B/\hbar, \label{cha}\\
\omega_{\alpha}(B)+\omega_{\beta}(B) &= -\frac{A(2I+1)}{2\hbar}\Bigg(\sqrt{1+\frac{2}{2I+1}x+x^{2}}+\sqrt{1-\frac{2}{2I+1}x+x^{2}}\Bigg).
\label{he}
\end{align}
According to Eq.~(\ref{cha}) and (\ref{he}), one can find that the value of the magnetic field can be replaced by the frequency difference between the two transitions. Furthermore, by substituting Eq.~(\ref{cha}) into Eq.~(\ref{he}), we can obtain an magnetic-field-calibration-free expression for the hyperfine constant $A$ as
\begin{equation}\label{freeconstant}
A=-2\hbar \omega_{0} \sqrt{\frac{1-M^{2}}{(2I+1)^{2}-M^{2}}},
\end{equation}
where $\omega_{0}=(\omega_{\alpha}+\omega_{\beta})/2$ and $M=\frac{g_J-g_I}{g_I}\frac{\omega_{\alpha}-\omega_{\beta}}{\omega_{\alpha}+\omega_{\beta}}$.

\section*{II. Experimental details and microwave pulse generation}

The degenerate Fermi gas of approximately $N\simeq 6\times10^{6}$ atoms of $^{40}\mathrm{K}$ is first prepared in the $\lvert 9/2,9/2\rangle$ hyperfine Zeeman state in a 1064-nm crossed optical dipole trap (ODT) at $T\simeq 0.3\,T_F$~\cite{Shan2026PRA}. At this stage, the trap intensity is set to $0.1I_0$, where $I_0$ denotes the reference intensity at the center of the ODT formed by two mutually incoherent linearly polarized 1064~nm laser beams propagating in orthogonal directions, with powers of $3.5\,\mathrm{W}$ and $2.5\,\mathrm{W}$ and waists of approximately 45\,$\mu\mathrm{m}$. Different trap intensities are obtained by proportional scaling of the laser powers at fixed beam waist.
For the magnetically insensitive tansitions, we transfer atoms into two Zeeman
states $|9/2, -1/2\rangle$ and $|9/2, +1/2\rangle$ via a rapid adiabatic passage driven by a radio-frequency (RF) transition. Starting from these states, we address two magnetically insensitive single-photon microwave (MW) transitions, $\alpha$ ($|9/2,-1/2\rangle$$\Leftrightarrow$$|7/2,+1/2\rangle$) and $\beta$ ($|9/2,+1/2\rangle$$\Leftrightarrow$$|7/2,-1/2\rangle$).
Ramsey spectroscopy is carried out with two square $\pi/2$ microwave pulses of duration $\tau_p=0.15\,\mathrm{ms}$, separated by a free-evolution time $T_R$. For state-resolved detection, we map atoms from the $F=7/2$ manifold onto well-separated Zeeman states in the $F=9/2$ manifold using an adiabatic Landau--Zener transfer. Specifically, atoms in $|7/2, +1/2\rangle$ are transferred to $|9/2, +3/2\rangle$, while atoms in $|7/2, -1/2\rangle$ are transferred to $|9/2, -3/2\rangle$. The resulting state populations are then determined by time-of-flight absorption imaging under a Stern--Gerlach magnetic-field gradient, which spatially separates the different Zeeman components.

The two magnetically insensitive transitions are driven by microwave fields under a homogeneous bias magnetic field $B$ applied along the $z$ direction, which defines the quantization axis. In our experiment, $B$ is varied between $0.7\mathrm{G}$ and $2.4~\mathrm{G}$. This field range is chosen to avoid spin-flip and spin-exchange processes at lower fields while keeping the residual sensitivity to magnetic-field fluctuations under control at higher fields~\cite{Peng2018CPL}. To further suppress magnetic-field noise, the programmable DC power supply (low-noise current source Delta ES030-5) is operated in manual constant-current mode, with the output voltage set only as high as needed to sustain the required current.

\begin{figure}[h!]
\includegraphics[width=6.0in]{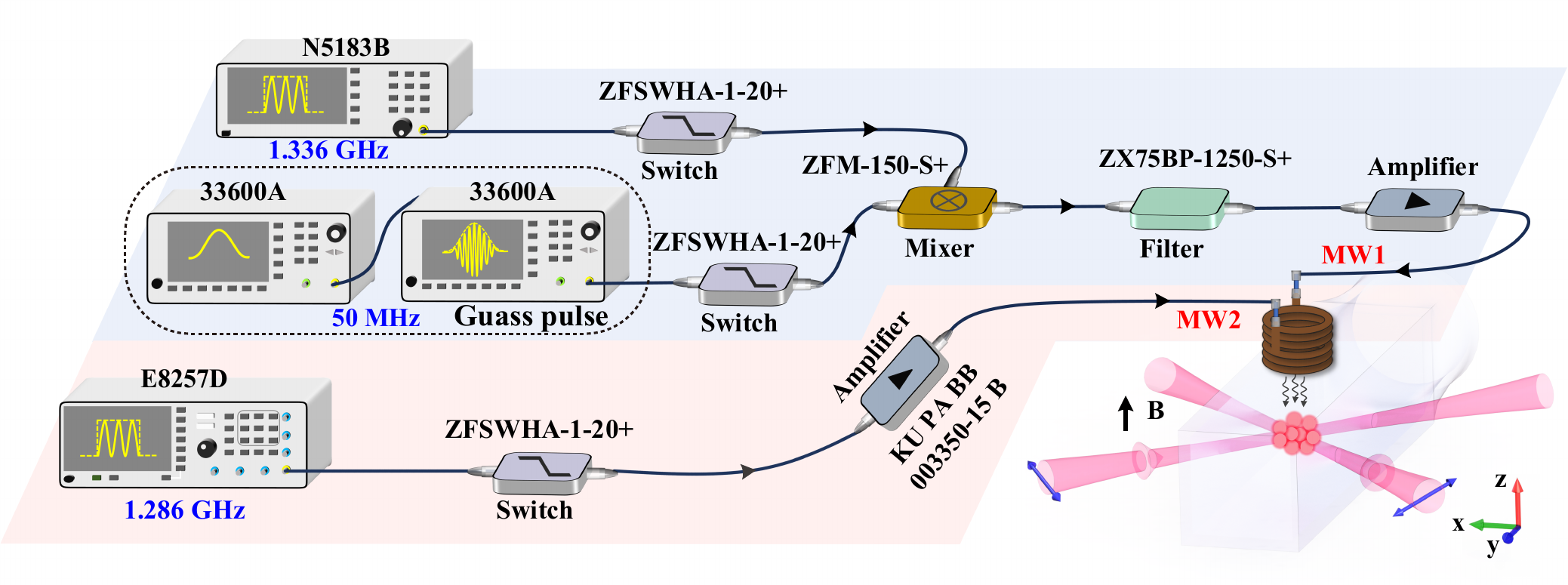}
\caption{Schematic of the microwave signal generation chain and its delivery to the atomic sample.
The upper branch (indicated by a light-blue shaded region) shows the Gaussian-shaped microwave (MW1) path, and the lower-left branch (indicated by a light-red shaded region) shows the square-pulse microwave (MW2) path.
The atoms are confined in a crossed optical dipole trap formed by two mutually incoherent linearly polarized 1064 nm laser beams propagating in orthogonal directions in the $xy$~plane.
A magnetic field $B$ applied along the $\hat{z}$ direction provides Zeeman splitting and defines the quantization axis.
}
\label{Microwave}
\end{figure}

Two microwave (MW) signal paths are employed in the experiment to enable both rapid coarse identification and high-precision determination of the resonance frequencies. The first microwave path (MW1), shown in the upper part of Fig.~\ref{Microwave}, employs Gaussian-shaped pulses for spectroscopy to suppress spectral sidelobes and thereby quickly locate the resonance at the 100-Hz level. The signal in this path is generated by mixing a 1.336~GHz carrier with a Gaussian-shaped radio-frequency sideband near 50~MHz~\cite{Ziliang2023CPB}, where the Gaussian envelope is provided by two external waveform generators. Under computer control, the first generator (Keysight 33600A) operates in arbitrary-waveform mode to output a Gaussian envelope, which is fed into the external modulation input of a second generator of the same model. The second generator continuously outputs a sinusoidal waveform with external modulation enabled, such that the Gaussian envelope modulates the sine-wave amplitude to produce a Gaussian-shaped radio-frequency pulse. Both the 1.336~GHz source and the 50~MHz Gaussian-envelope generator are phase-locked to a common 10~MHz reference, and the unwanted (positive) sideband is removed by a bandpass filter (ZX75BP-1250-S$+$), yielding a control signal centered at 1.286~GHz~\cite{Graham2022Nature}. The mixed signal is then amplified to a maximum output power of 5~W using a power amplifier (Mini-Circuits, ZHL-5W-2G-S$+$) and delivered to the atoms via the MW1 antenna along the vertical direction. The second microwave path (MW2), operating at 1.286~GHz and shown in the lower part of Fig.~\ref{Microwave}, provides square pulses for Ramsey interferometry, enabling Hz-level precision in the determination of the resonance frequency. The signal in this path is generated by a low-phase-noise microwave analog signal generator (Keysight E8257D), amplified to a maximum output power of 15~W using a low-noise power amplifier (KU PA BB 003350-15 B), and delivered to the atoms via the MW2 antenna along the same vertical direction.

\section*{III. Ramsey measurement and fitting}

The fitting model used for the Ramsey oscillation is given by~\cite{Ivannikov2018JPB, Xu2019RSI, Meinert2024PRX, Bourdel2025RSI}
\begin{equation}\label{Ramseyfitting}
\eta(T_R)=\eta_0+C_0\exp\left[-\frac{1}{2}\left(\frac{T_R}{T_\phi}\right)^2\right]\cos(\delta T_R+\phi_0),
\end{equation}
where $\eta(T_R)$ is the measured population imbalance after the Ramsey free-evolution time $T_R$.
The parameters $\eta_0$ and $C_0$ denote the offset and the overall Ramsey-fringe contrast, respectively.
The Gaussian envelope describes the contrast decay due to dephasing.
The coherence time $T_\phi$ is defined as the free-evolution time at which the root-mean-square accumulated phase uncertainty reaches one radian, i.e., $\sigma_\phi=\sigma_\omega T_\phi=1$, where $\sigma_\omega$ denotes the root-mean-square fluctuation of the transition angular frequency.
The cosine term describes the Ramsey oscillation as a function of the accumulated phase $\delta T_R+\phi_0$, where $\delta=\omega_{\mathrm{MW}}-\omega$ is the angular detuning between the applied microwave field $\omega_{\mathrm{MW}}$ and the atomic transition angular frequency $\omega$, and $\phi_0$ is a constant phase offset.

The transition angular frequencies are determined from $\omega_{\alpha,\beta}=\omega_{\mathrm{MW}}-\delta_{\alpha,\beta}$, where $\delta_{\alpha,\beta}$ is the fitted angular detuning for the $\alpha$ or $\beta$ transition. In our measurements, the two transitions are measured in an interleaved sequence: the $\alpha$ and $\beta$ transitions are probed alternately at identical free-evolution times. Each individual measurement at a given transition and free-evolution time takes about 2 minutes. To obtain a more reliable estimate of the frequency uncertainty, each data point is obtained by averaging three repeated measurements. The statistical standard error of these three measurements is used as the point-wise uncertainty in the global Ramsey fit. The frequency uncertainty is obtained from the fit covariance of the global Ramsey fit, with the repeated-measurement uncertainty at each data point incorporated. A complete set of Ramsey fringes for the $\alpha$ and $\beta$ transitions, consisting of about $2\times 3\times 160$ individual measurements, therefore takes about 32 hours to acquire.

The interleaved measurement makes the time interval between neighboring
$\alpha$ and $\beta$ measurements much shorter than the characteristic time
scale of slow magnetic-field drift. Therefore, for the slow-drift component of the magnetic-field noise, neighboring $\alpha$ and $\beta$ measurements can be regarded as experiencing approximately the same magnetic field. In this case, although slow magnetic-field drift can introduce correlations between the extracted values of the two transition angular frequencies, the magnetic-field dependence of the extracted hyperfine constant $A$ is completely eliminated by Eq.~(\ref{freeconstant}). This means that the correlated contribution associated with the common slow magnetic-field drift is eliminated in the determination of $A$, as expressed by
\begin{equation}\label{correlation}
\frac{\partial A}{\partial \omega_\alpha}\frac{\partial \omega_\alpha}{\partial B}
+
\frac{\partial A}{\partial \omega_\beta}\frac{\partial \omega_\beta}{\partial B}
=0 .
\end{equation}
In this sense, the interleaved measurement strongly suppresses the influence of slow magnetic-field drift. Nevertheless, during the acquisition of a full Ramsey fringe, slow magnetic-field drift can still lead to shot-to-shot phase wander within each Ramsey scan. This effect increases the scatter of the Ramsey data and is included in the fitted uncertainty of the transition angular frequency. It should be noted that the above cancellation only applies to the slow-drift component of the magnetic-field noise. High-frequency magnetic-field noise has a correlation time much shorter than the switching interval between the $\alpha$ and $\beta$ measurements, and therefore can be approximately regarded as uncorrelated. Since the magnetic-field noise contains both slow correlated and faster uncorrelated components, we conservatively propagate the magnetic-field-induced frequency uncertainties of the two transitions as independent contributions when estimating the uncertainty in the hyperfine constant $A$.

\section*{IV. Precise Determination of hyperfine constant $A$}

\subsection{A. Zero-intensity extrapolation of the AC Stark shift}

In our experiment, the ODT is formed by
far-detuned linearly polarized laser beams. In the low-intensity range, the dominant light shift arises from the
second-order AC Stark effect and is therefore expected
to scale linearly with optical intensity~\cite{DSteck}. This allows
us to extrapolate the resonance frequency to zero optical
intensity via a linear fit, thereby effectively eliminating
the trap-induced AC Stark frequency shift. In contrast, circularly polarized light can introduce a magnetic-field-dependent vector light shift and higher-order Stark effects, which together can lead to a nonlinear dependence on the optical intensity~\cite{Zhan2016PRL}. For each bias magnetic field, we measure $\omega_\alpha$ and $\omega_\beta$ as a function of optical intensities $I$ and fit the data with the linear model
\begin{equation}
\omega_{\alpha,\beta}(B,I)
=
\omega^0_{\alpha,\beta}(B)
+
k_{\alpha,\beta}(B)I,
\end{equation}
where the fit intercept gives the zero-intensity transition frequency $\omega^0_{\alpha,\beta}(B)$, and $k_{\alpha,\beta}(B)$ is the corresponding linear light shift coefficient.

To verify the validity of the linear model, we analyzed the data and compared the residuals of the linear fits with the corresponding measurement uncertainties. As summarized in Table~\ref{tab:linear_fit_residuals}, for selected data sets, the residuals are significantly smaller than the associated uncertainties, thereby quantitatively confirming the validity of the linear model. Moreover, the residual analysis shows no systematic nonlinear trend, even at relatively high magnetic fields, indicating that the linear model remains applicable in this regime.

\begin{table*}[h]
\caption{
Residual analysis of the linear fits and comparison with the associated measurement uncertainties.
}
\label{tab:linear_fit_residuals}

\begin{ruledtabular}
\begin{tabular}{cc|cccc|cccc}
$B$ &
$I/I_0$ &
\multicolumn{4}{c|}{$(\omega_\alpha-\omega_\alpha^0)/2\pi\ (\mathrm{Hz})$} &
\multicolumn{4}{c}{$(\omega_\beta-\omega_\beta^0)/2\pi\ (\mathrm{Hz})$} \\
$(\mathrm{G})$ &
&
measured &
fitted &
residuals &
uncertainties &
measured &
fitted &
residuals &
uncertainties \\
\hline
\multirow{3}{*}{0.9}
& 0.1 & -4.88  & -4.95  & 0.07  & 0.68 & -4.83  & -4.93  & 0.10  & 0.64 \\
& 0.3 & -15.02 & -14.85 & -0.17 & 0.79 & -15.02 & -14.79 & -0.23 & 0.69 \\
& 0.5 & -24.61 & -24.75 & 0.14  & 0.98 & -24.47 & -24.65 & 0.18  & 0.87 \\
\hline
\multirow{3}{*}{2.3}
& 0.1 & -6.94  & -6.09  & -0.85 & 1.85 & -7.89  & -6.95  & -0.94 & 1.81 \\
& 0.3 & -16.35 & -18.27 & 1.92  & 1.97 & -18.73 & -20.84 & 2.11  & 1.92 \\
& 0.5 & -31.75 & -30.44 & -1.31 & 2.29 & -36.12 & -34.74 & -1.38 & 2.19
\end{tabular}
\end{ruledtabular}
\end{table*}

\subsection{B. Final determination of $A$ by inverse-variance weighted averaging}

After determining the transition frequencies $\omega_\alpha$ and $\omega_\beta$, we obtain the hyperfine constant
$A$ according to Eq.~(\ref{freeconstant}).
For different bias magnetic fields, the measured transition frequencies and the corresponding values of $A$ are summarized in Table~\ref{tab:freq_A_different_B}, where the ODT-induced frequency shifts have been removed by zero-intensity extrapolation, as described in detail in Fig.~3 of the main text and Sec.~IV~A of the Supplemental Material. The final value of the hyperfine constant $A$ is then determined
using the inverse-variance weighted average,
\begin{equation}\label{hyperfineconstantA}
A=
\frac{\sum_i A_i/\sigma_{A,i}^2}{\sum_i 1/\sigma_{A,i}^2},
\end{equation}
where $A_i$ and $\sigma_{A,i}$ denote the hyperfine constant and its uncertainty obtained at the $i$th magnetic field, respectively (see Sec. V for a detailed uncertainty budget). This gives the final value of the hyperfine constant as
$A=-h\times285.730536~\mathrm{MHz}$.

\begin{table*}[htbp]
\centering
\caption{Transition frequencies and the corresponding hyperfine constant $A$ at different bias magnetic fields.}
\label{tab:freq_A_different_B}
\setlength{\tabcolsep}{8pt}
\begin{tabular}{c|cccccc}
\hline\hline
$B~(\mathrm{G})$
& $0.76$ & $0.96$ & $1.18$ & $1.45$ & $1.84$ & $2.35$ \\
\hline
$\omega_\alpha/2\pi~(\mathrm{MHz})$
& $1285.78933930$ & $1285.79048039$ & $1285.79192815$ & $1285.79411847$ & $1285.79806157$ & $1285.80469647$  \\
$\omega_\beta/2\pi~(\mathrm{MHz})$
& $1285.78896354$ & $1285.79000274$ & $1285.79134176$ & $1285.79340172$ & $1285.79715126$ & $1285.80353391$  \\
$A_i/h~(\mathrm{MHz})$
& $-285.730535$
& $-285.730538$
& $-285.730530$
& $-285.730536$
& $-285.730526$
& $-285.730537$ \\
\hline\hline
\end{tabular}
\end{table*}

\section{V. Uncertainty budget of hyperfine constant $A$}
\label{sec:uncertainty_budget}

In this section, we provide a comprehensive account of the uncertainty budget for the determination of the hyperfine constant $A$. In our experiment, measurements are performed at several different values of the bias magnetic field $B$. For each measurement at a given bias magnetic field $B$, we first estimate the contributions from all relevant systematic effects as
\begin{equation}\label{Allcontribution}
\sigma_{A,i}
=
\sqrt{
\sigma_{A,B}^{2}
+
\sigma_{A,\mathrm{AC}}^{2}
+
\sigma_{A,\rho}^{2}
+
\sigma_{A,g_I}^{2}
+
\sigma_{A,g_J}^{2}
},
\end{equation}
where $\sigma_{A,i}$ is the uncertainty in $A$ at the $i$-th magnetic field, and $\sigma_{A,B}$, $\sigma_{A,\mathrm{AC}}$, $\sigma_{A,\rho}$, $\sigma_{A,g_I}$ and $\sigma_{A,g_J}$ denote the respective uncertainty contributions from magnetic-field fluctuations, the zero-intensity extrapolation of the AC Stark shifts, density-dependent frequency shifts, and the uncertainties in the Land\'{e} $g$ factors $g_I$ and $g_J$.

\subsection{A. Magnetic-field fluctuations}

According to Eq.~(\ref{freeconstant}), the hyperfine constant $A$ is theoretically independent of the magnetic field $B$. However, in the experiment, the two transition angular frequencies, $\omega_\alpha$ and $\omega_\beta$, cannot be measured simultaneously. As a result, temporal fluctuations of the magnetic field propagate into the uncertainty in $A$ through fluctuations in the measured transition angular frequencies. The contribution of magnetic-field fluctuations to the uncertainty in $A$ can be estimated by error propagation as
\begin{equation}
\sigma_{A,B}=\sqrt{\left(\frac{\partial A}{\partial \omega_\alpha}\frac{\partial \omega_\alpha}{\partial B}\right)^2
+\left(\frac{\partial A}{\partial \omega_\beta}\frac{\partial \omega_\beta}{\partial B}\right)^2}\delta B,
\end{equation}
where the sensitivity coefficients
$\frac{\partial A}{\partial \omega_\alpha}$,
$\frac{\partial A}{\partial \omega_\beta}$,
$\frac{\partial \omega_\alpha}{\partial B}$ and
$\frac{\partial \omega_\beta}{\partial B}$
are calculated according to the Breit--Rabi formula. In the low magnetic-field regime of the present experiment, the $\alpha$ and $\beta$ transition angular frequencies can be approximated as
\begin{equation}\label{alphafrequency}
\omega_\alpha \approx -\frac{9}{2}A + \mu_B g_I B + C_2 B^2 + O(B^4),
\end{equation}
\begin{equation}\label{betafrequency}
\omega_\beta \approx -\frac{9}{2}A - \mu_B g_I B + C_2 B^2 + O(B^4),
\end{equation}
with the second-order Zeeman coefficient $C_2 = -\frac{80\mu_B^2(g_J-g_I)^2}{729A}$. For the magnetic-field range used here, $B = 0.7 \sim 2.4~\mathrm{G}$, the angular frequency variation induced by magnetic-field fluctuations is dominated by the second-order Zeeman term and can be approximated as
\begin{equation}\label{secondorderZeeman}
\frac{\partial \omega_\alpha}{\partial B}
\approx
\frac{\partial \omega_\beta}{\partial B}
\approx 2C_2B.
\end{equation}
The transition angular frequency uncertainty induced by magnetic-field noise can be expressed as
\begin{equation}\label{fluctuation1}
\sigma_{\omega_\beta}
=
\frac{\partial \omega_\beta}{\partial B}\delta B.
\end{equation}
According to the relation $\sigma_{\omega_{\beta}}=1/T_\phi$ established in Sec.~III, the transition angular frequency uncertainty and the corresponding magnetic field noise can be estimated from the fitted Ramsey coherence time $T_\phi$ as
\begin{equation}\label{fluctuation2}
\sigma_{\omega_\beta}=\frac{\partial \omega_\beta}{\partial B}\delta B=\frac{1}{T_\phi}.
\end{equation}
The resulting contributions of magnetic-field fluctuations to the uncertainty in $A$ at different bias magnetic fields are summarized in Table~\ref{tab:magnetic_field_sensitivity}.

\begin{table*}[h]
\caption{Magnetic-field fluctuations and their contribution to the uncertainty budget in $A$ at different bias magnetic fields.
}
\label{tab:magnetic_field_sensitivity}
\begin{ruledtabular}
\begin{tabular}{cccccccc}
$B$ &
$T_\phi$ &
$\delta B$ &
$\dfrac{\partial A}{\partial \omega_\alpha}/\hbar$ &
$\dfrac{\partial A}{\partial \omega_\beta}/\hbar$ &
$\dfrac{\partial \omega_\alpha}{\partial B}/2\pi$ &
$\dfrac{\partial \omega_\beta}{\partial B}/2\pi$ &
$\sigma_{A,B}/h$ \\
$(\mathrm{G})$ &
$(\mathrm{ms})$ &
$(\mu\mathrm{G})$ &
&
&
$(\mathrm{kHz}/\mathrm{G})$ &
$(\mathrm{kHz}/\mathrm{G})$ &
$(\mathrm{Hz})$ \\
\hline
0.76 & 378 & 97  & 1.95 & -2.17 & 4.83  & 4.34  & 1.29 \\
0.96 & 300 & 95  & 2.51 & -2.73 & 6.04  & 5.54  & 2.03 \\
1.18 & 270 & 93  & 3.11 & -3.33 & 7.36  & 6.87  & 3.01 \\
1.45 & 192 & 98  & 3.83 & -4.05 & 8.99  & 8.50  & 4.77 \\
1.84 & 143 & 102 & 4.88 & -5.10 & 11.35 & 10.85 & 7.98 \\
2.35 & 108 & 105 & 6.28 & -6.50 & 14.42 & 13.93 & 13.45
\end{tabular}
\end{ruledtabular}
\end{table*}

\subsection{B. AC Stark shift}

The AC Stark contribution to the uncertainty in $A$ arises primarily from the frequency uncertainties associated with the zero-intensity extrapolation. Since no resolvable common-mode correlation between the zero-intensity intercepts of the two transitions is identified, the AC Stark extrapolation uncertainties of the transition frequencies are propagated as independent contributions, and are evaluated by error propagation as
\begin{equation}\label{ACshift}
\sigma_{A,\mathrm{AC}}
=
\sqrt{
\left(
\frac{\partial A}{\partial \omega_\alpha}
\right)^2
\sigma_{\omega_\alpha,\mathrm{AC}}^2
+
\left(
\frac{\partial A}{\partial \omega_\beta}
\right)^2
\sigma_{\omega_\beta,\mathrm{AC}}^2
},
\end{equation}
where $\sigma_{\omega_\alpha,\mathrm{AC}}$ and $\sigma_{\omega_\beta,\mathrm{AC}}$ denote the angular frequency uncertainties arising from the zero-intensity extrapolation, which are obtained from the fit covariance of the linear extrapolation model, with the repeated-measurement uncertainty at each ODT intensity incorporated in the fit. The resulting AC Stark shift contributions to the uncertainty budget of the hyperfine constant $A$ at different bias magnetic fields are summarized in Table~\ref{tab:uncertainty_budget}.

\subsection{C. Density-dependent frequency shift}

The contribution of density-dependent shifts to the uncertainty in $A$ can be neglected. In our measurement, the atoms are initially prepared in a single spin-polarized degenerate Fermi gas and the microwave field coherently drives the transition to the other hyperfine state. Ideally, all atoms remain in the same spinor state during
the coherent interrogation. Therefore, due to the indistinguishability of identical fermions and the Pauli exclusion principle, the zero-range s-wave collisions are suppressed, and no conventional s-wave mean-field clock shift is expected, as discussed in Ref.~\cite{Ketterle2003Science}. More generally, for a coherently driven ultracold two-state Fermi gas, the microwave pulse performs a coherent rotation in the two-state Hilbert space. Since the fermionic contact interaction is invariant under this rotation, the microwave excitation does not change the interaction energy, and no interaction-induced spectroscopic shift is expected, as shown in Ref.~\cite{Zwierlein2003PRL}. Experimentally, this expectation is confirmed by the atomic density scan shown in Fig.~4 of the main text, which shows no systematic dependence of the extracted hyperfine constant $A$ on atomic density. The observed variations in $A$ are limited to random fluctuations at the sub-Hz level, well below
the measurement uncertainty. Although residual shifts associated with higher-partial-wave collisions may exist at the sub-Hz level, they cannot be reliably resolved under our present experimental conditions. We therefore treat the contribution of density-dependent shifts to the uncertainty in $A$ as negligible.

\subsection{D. Land\'{e} $g$-factor sensitivity}

The determination of the hyperfine constant $A$ in our scheme depends on the precise values of the electronic and nuclear Land\'{e} $g$ factors $g_J$ and $g_I$, as given by Eq.~(\ref{freeconstant}).
In our experiment, we use the most precise values currently available, namely $g_J=2.00229421(24)$ and $g_I=0.000176490(34)$\cite{Daniel2019K40,Arimondo1977RMP}.
To quantify how the uncertainties of these Land\'{e} factors propagate into the hyperfine constant $A$, for fixed experimental parameters we vary $g_I$ and $g_J$ independently within their quoted $1\sigma$ uncertainties and analyze the resulting variation in $A$.
Specifically, we write $g_k=g_k^{(0)}\pm \delta g_k$ with $k\in\{I,J\}$, where $g_k^{(0)}$ is the central value with $g_I^{(0)}=0.000176490$ and $g_J^{(0)}=2.00229421$, and $\delta g_k$ denotes the corresponding $1\sigma$ uncertainty with $\delta g_I=3.4\times10^{-8}$ and $\delta g_J=2.4\times10^{-7}$.
The resulting variations in $A$ due to the uncertainties in the Land\'{e}  $g$ factors $g_I$ and $g_J$ are shown in Fig.~\ref{Lande}(a) and Fig.~\ref{Lande}(b), respectively.
When $g_I$ and $g_J$ are scanned over their full $1\sigma$ ranges, the corresponding uncertainty contributions are found to be $\sigma_{A,g_I}\simeq2\pi\times0.15~\mathrm{Hz}$ and $\sigma_{A,g_J}\simeq2\pi\times0.00009~\mathrm{Hz}$, respectively.
Thus, the uncertainty contribution from the Land\'{e} factors is dominated by the nuclear Land\'{e} factor $g_I$.
Nevertheless, this contribution remains sufficiently small for the Hertz-level determination of $A$.
The contributions of the Land\'{e} $g$ factors $g_I$ and $g_J$ to the uncertainty budget of the hyperfine constant $A$ were further evaluated at different bias magnetic fields and are summarized in Table~\ref{tab:magnetic_field_sensitivity}.

\begin{figure}[htb]
\centering
\includegraphics[width=0.69\linewidth]{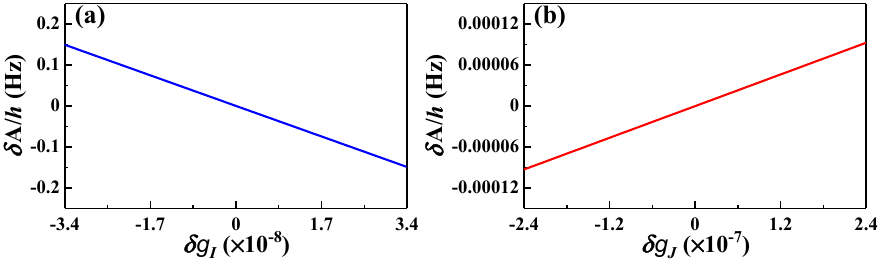}
\caption{Sensitivity of the hyperfine constant $A$ to the Land\'{e} $g$ factors.
Panel (a) shows the deviation of $A$ from its nominal value when $g_I$ is scanned within its quoted $1\sigma$ uncertainty, and panel (b) shows the corresponding result for $g_J$. For each panel, the Land\'{e} factor is varied as $g_k=g_k^{(0)}\pm \delta g_k$ with $k\in\{I,J\}$, where $g_k^{(0)}$ is the central value with $g_I^{(0)}=0.000176490$ and $g_J^{(0)}=2.00229421$ and $\delta g_k$ denotes the quoted $1\sigma$ uncertainty of $g_k$. The transition frequencies used in the evaluation are measured at an optical dipole trap intensity of $I=0.1I_0$ and a magnetic field of $B=0.76~\mathrm{G}$.}
\label{Lande}
\end{figure}

\subsection{E. Total uncertainty from inverse-variance weighted averaging}

The uncertainty contributions from different sources at various magnetic fields are summarized in Table~\ref{tab:uncertainty_budget} Finally, the uncertainties obtained at different magnetic fields are combined using an inverse-variance weighted average, and the total uncertainty of the measured hyperfine constant $A$ is given by
\begin{equation}\label{totaluncertainty}
\sigma_A
=
\left(
\sum_i \frac{1}{\sigma_{A,i}^2}
\right)^{-\frac{1}{2}},
\end{equation}
where $\sigma_{A,i}$ denotes the uncertainty of the hyperfine constant $A$ at the $i$th magnetic field. Applying this procedure to the data in Table~\ref{tab:uncertainty_budget} gives a final uncertainty of
$\sigma_A=h\times 2~\mathrm{Hz}$.

\begin{table*}[htbp]
\caption{
Uncertainty budget for the determination of the hyperfine constant $A$ at different bias magnetic fields.
}
\label{tab:uncertainty_budget}

\begin{ruledtabular}
\begin{tabular}{ccccccc}
$B$ &
$\sigma_{A,B}/h$ &
$\sigma_{A,\mathrm{AC}}/h$ &
$\sigma_{A,g_I}/h$ &
$\sigma_{A,g_J}/h$ &
$A_i/h$ &
$\sigma_{A,i}/h$ \\
$(\mathrm{G})$ &
$(\mathrm{Hz})$ &
$(\mathrm{Hz})$ &
$(\mathrm{Hz})$ &
$(\mathrm{Hz})$ &
$(\mathrm{MHz})$ &
$(\mathrm{Hz})$ \\
\hline
0.76 & 1.29  & 1.62  & 0.15 & 0.00009 & -285.730535 & 2.08  \\
0.96 & 2.03  & 1.11  & 0.24 & 0.00015 & -285.730538 & 2.33  \\
1.18 & 3.01  & 7.02  & 0.36 & 0.00023 & -285.730530 & 7.65  \\
1.45 & 4.77  & 9.43  & 0.54 & 0.00034 & -285.730536 & 10.58 \\
1.84 & 7.98  & 7.42  & 0.87 & 0.00054 & -285.730526 & 10.94 \\
2.35 & 13.45 & 26.70 & 1.43 & 0.00089 & -285.730537 & 29.93 \\
\end{tabular}
\end{ruledtabular}
\end{table*}

\end{widetext}

\bibliographystyle{apsrev4-2}
\bibliography{reference-Hyperfine-Constant}